\documentclass[preprint,
superscriptaddress,
amsmath,amssymb,
aps,
prb,
]{revtex4-2}
\usepackage[table,xcdraw]{xcolor}
\usepackage{graphics}
\usepackage{dcolumn}
\usepackage{bm}
\usepackage{array}
\usepackage{booktabs}
\usepackage{dsfont}
\usepackage{multirow}
\usepackage{esint}
\usepackage{color}
\usepackage{graphicx}
\usepackage{epstopdf}
\usepackage{verbatim}
\usepackage{silence}
\usepackage{float}
\usepackage{fancyvrb}
\usepackage[export]{adjustbox}
\usepackage{svg} 
\usepackage[normalem]{ulem}

\usepackage{tikz}
\usepackage{pgfplots}
\usepackage{pgfplotstable}
\pgfplotsset{compat = newest}
\usetikzlibrary{shapes.geometric, arrows}

\usepackage{scalerel}
\usepackage{tikz}
\usetikzlibrary{svg.path}
\definecolor{orcidlogocol}{HTML}{A6CE39}
\tikzset{
  orcidlogo/.pic={
    \fill[orcidlogocol] svg{M256,128c0,70.7-57.3,128-128,128C57.3,256,0,198.7,0,128C0,57.3,57.3,0,128,0C198.7,0,256,57.3,256,128z};
    \fill[white] svg{M86.3,186.2H70.9V79.1h15.4v48.4V186.2z}
                 svg{M108.9,79.1h41.6c39.6,0,57,28.3,57,53.6c0,27.5-21.5,53.6-56.8,53.6h-41.8V79.1z M124.3,172.4h24.5c34.9,0,42.9-26.5,42.9-39.7c0-21.5-13.7-39.7-43.7-39.7h-23.7V172.4z}
                 svg{M88.7,56.8c0,5.5-4.5,10.1-10.1,10.1c-5.6,0-10.1-4.6-10.1-10.1c0-5.6,4.5-10.1,10.1-10.1C84.2,46.7,88.7,51.3,88.7,56.8z};
  }
}
\newcommand\orcidicon[1]{\href{https://orcid.org/#1}{\mbox{\scalerel*{
\begin{tikzpicture}[yscale=-1,transform shape]
\pic{orcidlogo};
\end{tikzpicture}
}{|}}}} 

\newcommand{\bfk}{{\bf k}}

\newcommand{\beq}{\begin{equation}}
\newcommand{\eeq}{\end{equation}}
\newcommand{\beqs}{\begin{eqnarray}}
\newcommand{\eeqs}{\end{eqnarray}}
\newcommand{\beql}{\begin{equation} \label}

\let\oldFootnote\footnote
\newcommand\nextToken\relax

\renewcommand\footnote[1]{%
    \oldFootnote{#1}\futurelet\nextToken\isFootnote}

\newcommand\isFootnote{%
    \ifx\footnote\nextToken\textsuperscript{,}\fi}

\usepackage{amsmath}
\numberwithin{equation}{section}

\usepackage{silence}
\usepackage[colorlinks=true,
            linkcolor=blue,
            citecolor=red,
            urlcolor=cyan]{hyperref}
            
\begin{document}
\title{Bloch Oscillations and Landau-Zener Transitions in Flat-Band Lattices with Quadratic and Linear Band Touchings}
\author{Chenhaoyue Wang \orcidicon{0000-0000-0000-0000}}
\affiliation{Department of Materials Science and Engineering, University of California, Los Angeles, CA 90095, USA}
\author{Carlos J. Garc{\'\i}a-Cervera \orcidicon{0000-0002-8880-2782}}
\affiliation{Department of Mathematics, University of California, Santa Barbara, CA 93106, USA}
\affiliation{BCAM - Basque Center for Applied Mathematics, Bilbao, E-48009, Basque Country, Spain}
\author{Amartya S. Banerjee \orcidicon{0000-0001-5916-9167}}
\email{asbanerjee@ucla.edu}
\affiliation{Department of Materials Science and Engineering, University of California, Los Angeles, CA 90095, USA}
\date{\today}

\begin{abstract}
Bloch oscillations (BOs) describe the coherent oscillatory motion of electrons in a periodic lattice under a constant external electric field. Deviations from pure harmonic wave packet motion or \emph{irregular Bloch oscillations} can occur due to Zener tunneling (Landau-Zener Transitions or LZTs), with oscillation frequencies closely tied to interband coupling strengths. Motivated by the interplay between flat-band physics and interband coupling in generating irregular BOs, here we investigate these oscillations in Lieb and Kagome lattices using two complementary approaches: coherent transport simulations and scattering matrix analysis. 
In the presence of unavoidable band touchings, half-fundamental and fundamental BO frequencies are observed in Lieb and Kagome lattices, respectively --- a behavior directly linked to their distinct band structures. When avoided band touchings are introduced, distinct BO frequency responses to coupling parameters in each lattice are observed. Scattering matrix analysis reveals strong coupling and potential LZTs between dispersive bands and the flat band in Kagome lattices, with the quadratic band touching enhancing interband interactions and resulting in BO dynamics that is distinct from systems with linear crossings. In contrast, the Lieb lattice --- a three level system --- shows independent coupling between the flat band and two dispersive bands, without direct LZTs occurring between the two dispersive bands themselves. Finally, to obtain a unifying perspective on these results, we examine BOs during a strain-induced transition from Kagome to Lieb lattices, and link the evolution of irregular BO frequencies to changes in band connectivity and interband coupling.
\end{abstract}

\maketitle
\newpage 

\section{Introduction}
\label{sec:Introduction}
Bloch oscillations (BOs), a fundamental quantum mechanical phenomenon first predicted by Felix Bloch in 1929 \cite{bloch_uber_1929} and further explored by Zener \cite{zener1934theory} in his studies of dielectric breakdown, describe the periodic oscillatory motion of wave packets in the momentum space of a periodic potential when subjected to a constant external force field. Rather than accelerating continuously, electrons in crystalline solids, whose motion is governed by Bloch waves modulated by the lattice potential, experience a periodic oscillation of their wave vector \cite{zener1934theory}. This oscillatory behavior arises because the periodic potential introduces Bragg reflection upon the electron reaching the Brillouin zone boundary, thereby reversing its wavevector \cite{ziman_principles_1972}. The energy spectrum associated with this oscillatory motion forms a discrete set of equally spaced levels known as the Wannier-Stark ladder \cite{wannier1960effective, mendez1993wannier}, which provides the spectral signature of Bloch oscillations in the presence of a uniform electric field. Experimentally, BOs are  readily observed in semiconductor superlattices and ultracold atomic systems \cite{LEO1992943, PhysRevLett.76.4508, haller2010inducing}, due to the reduced scattering and longer coherence lengths in these systems. More recently, they have also been observed in superconducting quantum hardware platforms \cite{guo2021observation}. Such observations help solidify the importance of BOs as a fundamental phenomenon in condensed matter systems, with implications for high-speed electronics \cite{tans1997individual} and optoelectronic device design \cite{haug2009quantum}. While understanding single-band BOs offers insights into electron dynamics within periodic potentials, current research increasingly focuses on how interband coupling influences the frequencies of these oscillations, thereby deepening our understanding of quantum transport \cite{raizen1997new, zhang2010directed, sun_observation_2018, dittrich1998quantum}. 

In materials with unavoidable band crossings, such as those with linearly dispersive Dirac cones, deviations from regular BO frequencies can arise due to alterations in the wave packet periodicity \cite{PhysRevLett.108.175303}. In contrast, Landau-Zener transitions (LZTs) occur in regions where band crossings are avoided. Here, a small band gap opening near the avoided crossing of strongly coupled energy levels can transform an adiabatic evolution into a non-adiabatic process, inducing transitions between the energy branches \cite{franco_de_carvalho_nonadiabatic_2015}. These transitions originate from the quantum superposition of states near the avoided crossing, with the LZT probability determined by the interband coupling strength and the rate of energy level change \cite{PhysRevA.86.063613}. For example, Krueckl et al. \cite{krueckl_bloch-zener_2012} utilized the scattering matrix formalism to analyze the accumulation of the tunneling phase in graphene nanoribbons, a prototypical Dirac semimetal. This work provided a theoretical framework to explain deviations in BO frequencies due to LZTs and highlighted the role of Bloch-Zener oscillations in shaping current dynamics.

Materials with dispersionless electronic states, or \emph{flat bands}, represent an important frontier for exotic phenomena in condensed matter systems. They often host non-trivial topological states arising from strong electron correlations and lattice geometry \cite{Kane_2005, PhysRevX.4.011010, barreteau2017bird, sharma2025strain}. The quenched kinetic energy in flat bands enhances electron-electron interactions, while stable, flat-band edge states reduce scattering, thereby facilitating efficient transport over extended distances \cite{PhysRevX.4.011010, balents2020superconductivity, iglovikov2014superconducting}. Localization effects in these systems can lead to prolonged coherence times, potentially introducing novel phenomena under the LZT process, such as electric field-driven BOs in the presence of strong correlations \cite{Aidelsburger_2013}. 

Furthermore, the structural symmetries of flat-band lattices are intrinsically linked to their topological properties, suggesting potential applications in optoelectronic and quantum transport devices \cite{Liu2014FlatBands, sharma2025strain}. Numerical investigations into non-adiabatic transitions at quadratic band touchings and their impact on unconventional Bloch oscillations hold promise for advancing tunable quantum transport applications \cite{PhysRevLett.116.245301}. Irregular Bloch-Zener oscillations have been theoretically investigated in two-dimensional flat-band Dirac systems, such as the $\alpha$-T$_3$ lattice, using the adiabatic-impulse model to capture the dynamical evolution of phase across multiple time intervals \cite{ye_irregular_2023}. This study demonstrates that the presence of small energy gaps near Dirac cones enhances the occurrence of Landau-Zener transitions, leading to complex transition probabilities distributed in momentum space, which in turn produce intricate morphological patterns and nonsmooth temporal current responses. However, the emergence and characteristics of such irregular oscillations in other flat-band materials, such as the Lieb and Kagome lattices, which possess distinct band topologies and flat-band connectivity, remain an area requiring further exploration.

The Lieb and Kagome lattices, despite their distinct real-space geometries, share fundamental topological characteristics and can be interconnected through specific geometric transformations \cite{lim_dirac_2020}. Both lattices exhibit Dirac points and host flat bands, making them ideal model systems for investigating the impact of LZTs on transport phenomena in novel flat-band materials. The Kagome lattice, composed of corner-sharing triangles, displays a unique electronic structure featuring flat bands, Dirac nodes, and Van Hove singularities \cite{PhysRevB.80.113102}. Its geometric frustration leads to destructive interference, resulting in localized electron states and momentum-independent dispersion \cite{BERGHOLTZ_2013}, which are implicated in emergent phenomena, such as superconductivity and high-temperature ferromagnetism \cite{Yin_2022}. The Lieb lattice, while possessing a different structural arrangement, shares similar flat-band physics and topological invariants with the Kagome lattice \cite{Leykam_2018}. However, in the Lieb lattice, the flat band intersects a pair of Dirac points rather than being energetically isolated. Under the application of biaxial strain, a geometric transition from the Kagome to the Lieb lattice can occur, breaking rotational symmetry while preserving crucial mirror and inversion symmetries, thus maintaining key topological properties \cite{lim_dirac_2020}. This topological correspondence across the structure transformation has been corroborated by multiple theoretical approaches, including hybrid Wannier function analysis, Berry curvature mapping, and Chern number calculations \cite{jiang_topological_2019}.

Studies on one-dimensional systems with linear band crossings have provided fundamental insights into the analysis of scattering matrices during band coupling transitions and the characterization of transport behavior under external fields \cite{Ando_2005}\cite{Kane_2005}\cite{BEC_2006}. These findings can often be generalized to higher dimensional systems along high-symmetry paths within the Brillouin zone, offering a broader understanding of electronic transport in complex lattice structures. In this work, we employ both coherent transport model and scattering matrix approach to investigate the effect of LZT on Bloch oscillation frequencies in systems with band crossings. Our numerical approach involves simulating the time evolution of current density using a tight-binding Hamiltonian with periodic potential modulations under an applied electric field. Concurrently, theoretical predictions are obtained by analyzing the phase evolution based on the scattering matrix, constructed through a Taylor expansion of the Hamiltonian near the band-touching point. This dual approach enables a direct comparison between coherent transport theory and scattering matrix approach. The validity of these two approaches is first established by their application to the well-understood graphene system. We then extend these methods to explore the relationship between irregular Bloch oscillations and tunneling parameters in the novel flat-band materials, Kagome and Lieb lattices. These two lattices share topological equivalence under certain symmetry transformations, yet they differ in their band connectivity and dispersion, especially under strain. While the precise impact of this equivalence on LZT-modulated Bloch oscillations in non-adiabatic regimes is not fully resolved, it opens a promising direction for exploring dynamic control of BOs. In particular, the strain-driven geometric transition in the Kagome lattice, from a quadratic band touching to a pair of Dirac points with a flat band intersecting their center, as observed in the Lieb lattice, provides a unique platform to explore LZT effects and their potential implications for tunable quantum transport.

\section{Methodology for studying Bloch oscillations }
Graphene, with its characteristic linear band crossings in its electronic band structure and well-documented Bloch oscillations arising from these features, provides an ideal benchmark system for validating two distinct theoretical methodologies employed in this study: the coherent transport framework and the scattering matrix model. Both approaches offer effective tools for investigating the dynamics of Bloch oscillations, particularly in systems exhibiting complex band structures.

\subsection{Coherent transport framework}
Bloch oscillations are conventionally described through the motion of a single electron within a single energy band, an approximation often valid for metallic systems under weak electric fields. However, this standard approach typically does not directly yield experimentally observable macroscopic quantities, such as the transient current, and its extension to multiband systems, where interband transitions become significant, poses considerable challenges \cite{rosenstein_ballistic_2010}. To overcome these limitations, we adopt the coherent transport framework recently analyzed in the context of the quantum Hall effect by Canc{\`e}s {\it et al.} \cite{cances_coherent_2021}. This framework offers a robust method for directly calculating macroscopic observables, specifically the transient current density induced by an externally applied electric field, by considering the coherent time evolution of electronic states across the entire system at equilibrium.

The coherent transport framework begins with an arbitrary time-independent Hamiltonian $\mathcal{H}_{\bfk}$ defined in the $\bfk$ space and provides a prediction of the current density for the entire system at equilibrium through the time evolution of states. By virtue of gauge transformations, a uniform electric field {$\epsilon$} applied along the \bm{$e_{\beta}$}-direction can be converted into a linear-in-time magnetic potential, which leads to the wave-vector changing linearly with time, so the time-dependent Hamiltonian can be expressed  as $\mathcal{H}_{\bfk -\epsilon \bm{e_{\beta}}t}$. With the form of the Hamiltonian determined, the time evolution operator $U^{\epsilon}_{\beta, {\bfk}}(t)$ can be calculated through the Schr\"odinger equation via the following system of ordinary differential equations (ODEs):
\begin{equation}
    i\partial_tU^{\epsilon}_{\beta, \bfk}(t)=\mathcal{H}_{\bfk-\epsilon e_{\beta}t}U^{\epsilon}_{\beta, \bfk}(t),\ U^{\epsilon}_{\beta, \bfk}(0) = \mathds{1},
\label{coherent}
\end{equation}
where $\mathds{1}$ is the identity operator. During this study, we work in natural units with $\hbar=1$, so time is measured in $\mathrm{eV}^{-1}$. Sampling in $\bfk$ space during the time evolution was performed under a \( 300 \times 300 \) two-dimensional Monkhorst–Pack grid \cite{PhysRevB.13.5188}. Convergence with respect to the $k$-point mesh was verified for all studied lattices by confirming that the steady current density is preserved at the long-time limit of $t=50$. To avoid long-term drift caused by numerical errors, a second-order symplectic numerical method \cite{yoshida_construction_1990} was used for system (\ref{coherent}), instead of the Runge–Kutta–Fehlberg method (RK45) \cite{Fehlberg1969LoworderCR}. 

\begin{figure}[htbp]
\centering
\includegraphics[width=1.0\linewidth]{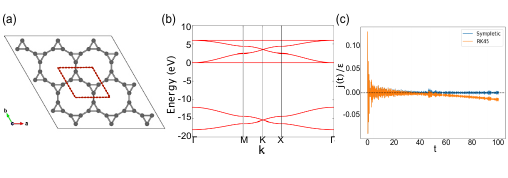}
\caption{(a) The top view of \( 3 \times 3 \times 1 \) superlattice structure, (b) corresponding band diagram, and (c) numerical current evolution over an extended time range computed using RK45 and symplectic ODE solvers, for the Kagome graphene lattice.}
\label{fig:coherent transport framework}
\end{figure}

As shown in Figure \ref{fig:coherent transport framework}b, when the Fermi level in the Kagome graphene lattice resides within the energy gap separating two isolated bands, the system exhibits insulating behavior, resulting in a vanishing current density in the long-time limit. Figure \ref{fig:coherent transport framework}c illustrates a comparison of current evaluations obtained using both the RK45 and the symplectic integrator. Over short time scales, both numerical methods yield current evolutions that are consistent with the expected insulating behavior. However, at longer time durations, the RK45 solution shows significant deviation from this expectation, with the calculated current drifting to non-physical negative values. In contrast, the second-order symplectic integrator maintains the expected absolute zero current density, even over extended simulation periods. Consequently, due to its superior long-time stability and preservation of physical constraints, the second-order symplectic numerical method is adopted for solving the time-dependent Schrödinger equation in all our numerical calculations. The macroscopic observable of interest in this study is the electric current density along a specific direction \bm{$e_{\alpha}$}, which is proportional to the expectation value of the velocity operator projected along the same direction $\bm{v}_{\alpha,\bfk}(t) = \partial_{k_{\alpha}}\mathcal{H}_{\bfk-\epsilon e_{\beta}t}$. For a multi-band system at equilibrium, the ground-state density matrix is given by
\begin{equation}
    \gamma_\bfk(0) = \mathds{1} (\mathcal{H}_\bfk\leqslant \mu_F) = \sum_{n:\ \epsilon_n(\bfk) \le \mu_F} |u_{n,\bfk}\rangle \langle u_{n,\bfk}|,
\end{equation}
where $\mu_F$ is the Fermi level, $n$ is the band index, $u_{n,\bfk}$ are the Bloch-functions associated to $\mathcal{H}(0)$, and $\epsilon_n(\bfk)$ are the corresponding energy levels.  This equation sets the initial density matrix $\gamma_\bfk(0)$ at $t = 0$, and implies that all electronic states with energies below $\mu_F$ are fully occupied, while those above $\mu_F$ are unoccupied. Then the density matrix at  time $t$, under the influence of an applied electric field, is obtained by evolving the initial density matrix using the time evolution operator:
\begin{equation}
    \gamma_{\beta,\bfk}^{\epsilon}(t) = U^{\epsilon}_{\beta, \bfk}(t)\gamma_\bfk(0)U^{\epsilon}_{\beta, \bfk}(t)^{\dagger}.
\end{equation}
where ${U^{\epsilon}_{\beta, \bfk}}(t)$ is the time evolution operator for a Bloch state with initial wavevector $\bfk$ subjected to an electric field $\epsilon$ applied along the direction $\bm{e_{\beta}}$. The macroscopic electric current density along a direction $\bm{e_{\alpha}}$ as a function of time is subsequently calculated by taking the trace of the velocity operator as applied to the time-evolved density matrix over all occupied Bloch states in the Brillouin zone:
\begin{equation}
    j_{\alpha,\beta}^{\epsilon}(t) = -(2\pi)^{-d}\int_{\mathcal{B}}\mathrm{Tr}[v_{\alpha,\bfk}(t)\gamma_{\beta,\bfk}^{\epsilon}(t)] d\bfk,
\end{equation}
where the integral is over the Brillouin zone $\mathcal{B}$ and $d$ is the dimension of the Brillouin zone. As described, this framework is independent of the specifics of the Hamiltonian, including factors like the number of bands and the spatial dimensions of the system. Provided the Hamiltonian is known and can be expressed analytically in the $\bfk$-space, coherent transport can be computed numerically. It is particularly well-suited for analyzing linear response regimes and adiabatic processes. The direct time evolution approach allows for effective evaluation of systems under various static fields, including direct current fields, and its ability to maintain equilibrium current over extended durations makes it ideal for capturing long-term phenomena such as Bloch oscillations and static conductivities, thus providing a powerful tool for investigating Bloch oscillations in complex multi-band systems. In all cases, the externally applied constant electric field is set to 0.1 in dimensionless units, where the field is scaled by a characteristic energy scale divided by a characteristic length scale of the specific system under investigation. A sufficiently long simulation time 4000 and a small electric field strength $\epsilon$ value of $1 \times 10^{-6}$ in the same dimensionless units are used to ensure accurate observation of Bloch oscillation phenomena without inducing significant non-linear effects. For all system models considered in this study, the Fermi level is consistently set to a value that results in a fully occupied lowest energy band and unoccupied higher energy bands at equilibrium, establishing a uniform basis for comparing the observed Bloch oscillation behavior under these specific initial conditions.

\begin{figure}[htbp]
\centering
\includegraphics[width=1.0\linewidth]{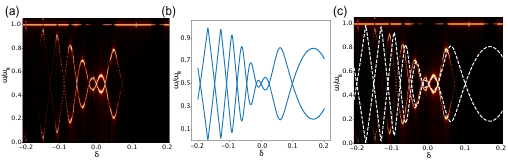}
\caption{Bloch oscillation frequencies as a function of the coupling parameter $\delta$ in the graphene system (see Sec.~\ref{subsec:scattering_matrix}). The frequencies correspond to oscillations in the current density: (a) calculated using the coherent transport framework, (b) obtained from the linearly expanded tight binding model, and (c) comparison between theoretical predictions and numerical results.}
\label{fig:BO_freq_graphene}
\end{figure}

Following the application of the coherent transport framework to the graphene system, specifically along the path encompassing the pair of symmetry-related Dirac points, a Fourier transformation of the calculated time-dependent current density is performed. The resulting frequency spectrum reveals a distinct peak corresponding to the numerically evaluated Bloch oscillation frequency, as presented in Figure \ref{fig:BO_freq_graphene}a. During an adiabatic process of a linear band crossing without Landau-Zener tunneling, as observed in graphene, the electronic wavefunction evolves as a coherent superposition of states from both bands. Because the electron traverses the Brillouin zone and returns to its initial state only after completing two full cycles through the linear crossing, the resulting Bloch oscillation exhibits a period twice that of the conventional case. This doubled periodicity directly translates to a Bloch oscillation frequency that is precisely half the fundamental Bloch frequency. When a small bandgap opens at the linear band crossing, forming an avoided band touching, interband tunneling becomes a non-negligible process in the region of closest energy approach. The quantum coherence established between the two interacting energy branches due to this tunneling gives rise to  LZT, which manifests itself as additional, irregular Bloch oscillation frequencies in the dynamical response of the system. These emergent frequencies exhibit a distinctive interweaving pattern around half the fundamental Bloch frequency, with the precise spectral positions and characteristics of this pattern exhibiting strong sensitivity to the tunneling strength parameter $\delta$, which is related to the probability of interband transitions. The observed asymmetric periodicity of the interweaving pattern, corresponding to positive and negative $\delta$ values, is a direct consequence of the asymmetry of the electronic band structure of graphene, specifically at the K point in the two-dimensional graphene Brillouin zone. This band structure asymmetry leads to different dynamical responses of the electronic states near the Dirac points when subjected to the external electric field and the resulting interband tunneling, ultimately shaping the observed irregular Bloch oscillation frequencies.

\subsection{Theoretical model: scattering matrix}
\label{subsec:scattering_matrix}
The LZT has been quantitatively analyzed in one-dimensional Dirac semimetals, such as graphene nanoribbons \cite{fahimniya_synchronizing_2021}. Furthermore, the impact of LZT on Bloch oscillations has been theoretically explored by directly deriving the scattering matrix of a tight-binding model under a constant electric field \cite{krueckl_bloch-zener_2012}. By exploiting the BZ symmetry in two-dimensional materials, the Bloch oscillations of graphene can be effectively reduced to a one-dimensional problem, facilitating the systematic study of similar patterns.

To describe graphene nanoribbons under a constant driving field, a one-dimensional Hamiltonian representing a gapped Dirac node is formulated as follows:
\begin{equation}
\mathcal{H} = \frac{2v\hbar}{a}\sin \left (\frac{a k(t)}{2}\right )\sigma_z+g\sigma_x,
\end{equation}
where $\sigma_z$ and $\sigma_x$ are the Pauli matrices acting on the sublattice degree of freedom, $g$ represents half of the band gap, $a$ is the lattice constant, and $v$ is the Fermi velocity.

The linear expansion of this time-dependent Hamiltonian at the Dirac point yields a form analogous to the prototypical LZT Hamiltonian. As a result, an LZT scattering matrix \cite{krueckl_bloch-zener_2012}, characterized by a tunneling probability and a phase acquired during tunneling, can be employed to describe the transitions between the electron and hole branches in this system. By analytically determining the eigenvalues of this scattering matrix, the dependence of the two irregular Bloch oscillation frequencies, arising from interband tunneling, on the applied electric field strength can be investigated.

The graphene system, as a two-dimensional Dirac semimetal, can be described by the following nearest-neighbor tight-binding model Hamiltonian:
\begin{equation}
\label{graphene_Hamiltonian}
\mathcal{H} = \begin{bmatrix}
0 & t_1\left(e^{i\bm{k}\cdot\bm{r}_1}+e^{i\bm{k}\cdot\bm{r}_2}+e^{i\bm{k}\cdot\bm{r}_3}\right) \\
t_1\left(e^{-i\bm{k}\cdot\bm{r}_1}+e^{-i\bm{k}\cdot\bm{r}_2}+e^{-i\bm{k}\cdot\bm{r}_3}\right) & 0
\end{bmatrix},
\end{equation}
where $t_1$ is the hopping energy, $\bm{r}_1 = \frac{a}{\sqrt{3}}(0,1)$, $\bm{r}_2 = \frac{a}{\sqrt{3}}\left(\frac{\sqrt{3}}{2},-\frac{1}{2}\right)$, $\bm{r}_3 = \frac{a}{\sqrt{3}}\left(-\frac{\sqrt{3}}{2},-\frac{1}{2}\right)$ and $a$ is the lattice constant. An electric field is applied along the direction of $\bm{b}_1 = \frac{2\pi}{a}\left(1,\frac{\sqrt{3}}{3}\right)$.

While graphene is inherently two-dimensional, the analysis of Bloch oscillations under a uniform electric field can be simplified to a quasi-one-dimensional problem by considering specific paths in the $k$ space. We focus on paths defined by
$\bm{k} = \alpha \bm{b}_1 + \left(\frac{1}{3}+\delta\right)\bm{b}_2,\ -0.5 \leqslant \alpha \leqslant 0.5$, where $\bm{b}_1 = \frac{2\pi}{a}\left(1,\frac{\sqrt{3}}{3}\right)$, $\bm{b}_2 = \frac{2\pi}{a}\left(0,\frac{2\sqrt{3}}{3}\right)$ and $\delta$ is a small parameter used to vary the path in the vicinity of the high-symmetry K point. As $\delta$ approaches zero, the charge carriers are driven by an external field to traverse a pair of Dirac cones in the BZ.

In this case, a simplified Hamiltonian can be obtained by performing a linear expansion around the crossing point. Focusing on the two lowest energy levels, the Hamiltonian describing the subspace they form is defined as:
\begin{equation}
\label{expansion_graphene}
    \mathcal{H_{\text{expand}}} = \pi t_1
    \begin{bmatrix}
        2\left(\alpha-\frac{1}{3}+\frac{\delta}{2}\right) & \sqrt{3}\delta \\
        \sqrt{3}\delta & -2\left(\alpha-\frac{1}{3}+\frac{\delta}{2}\right)
    \end{bmatrix}.
\end{equation}
For each considered path near $K$, the graphene system can be treated as a graphene nanoribbon \cite{krueckl_bloch-zener_2012} with a corresponding energy difference, $2g = 2\sqrt{3}\pi\delta t_1$, between the two branches. This energy difference, 2g, directly proportional to $\delta$, characterizes the effective coupling strength that governs the interband tunneling probability and consequently influences the resulting irregular Bloch oscillation frequencies by modifying the effective energy landscape experienced by the charge carriers. This mapping allows the application of the LZT scattering matrix, characterized by well-defined tunneling rates and phase accumulation, derived for the one-dimensional gapped Dirac system, to evaluate the irregular Bloch oscillation frequencies observed in the two-dimensional graphene system under an electric field driving carriers near the Dirac points. These frequencies are modulated by the effective energy gap that open at the linear crossing points in the graphene system, as illustrated in Figure \ref{fig:BO_freq_graphene}b.

We observe discrepancies between the numerical and theoretical results  when the  absolute value of $\delta$ exceeds a limit of 0.05. This deviation arises because, at larger $\delta$ values, the off-diagonal terms in the Hamiltonian become comparable in magnitude to the diagonal terms. As a result, the carrier motion across the Brillouin zone can no longer be treated as a scattering process from $t=-\infty$ to $t=+\infty$,  leading to a breakdown of the assumptions underlying the simple LZT scattering matrix formalism. However, for cases involving small effective energy bandgaps, cross-verification of the BO frequency dependence on the tunneling parameter in the graphene system, between the theoretical scattering matrix predictions and the numerical simulations based on the coherent transport framework, confirms the accuracy of the predictions made by both approaches. This successful validation establishes a foundation for extending the discussion of Bloch oscillations and LZT phenomena to more complex multiband systems, such as the three-energy-level Lieb and Kagome lattices, and paves the way for future dynamic studies of Bloch oscillations during strain-induced transformations between these lattice structures.

\section{Quadratic touching: unstrained Kagome system}
\begin{figure}[htbp] 
\centering 
\includegraphics[width=0.6\linewidth]{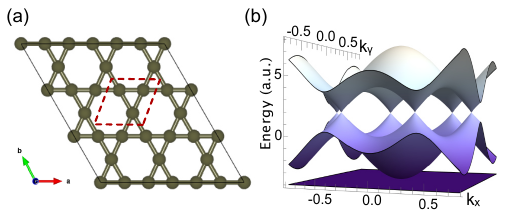} 
\caption{(a) Geometric structure of the Kagome lattice, with the red dashed line marking the unit cell, and (b) its electronic band diagram.} 
\label{Kagome_structure} 
\end{figure}

\begin{figure}[htbp]
\centering
\includegraphics[width=1.0\linewidth]{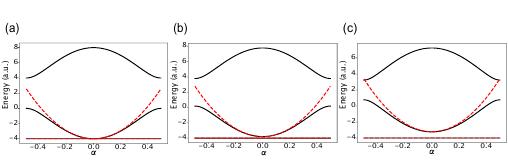}
\caption{ (a-c) Band structure of the Kagome lattice calculated from Equation~\ref{path} (black lines) and Equation~\ref{expansion} (red dashed lines) for coupling parameter $\delta$ = 0, 0.1 and 0.2, respectively.}
\label{fig:Kagome_band}
\end{figure}

The Kagome lattice, a three-level energy system characterized by a quadratic band touching point at the $\Gamma$ point, is constructed from two interconnected triangular structures within a single unit cell, as depicted in the top view in Figure \ref{Kagome_structure}a. Geometric deformation modifies the hopping interactions between atomic sites while preserving the quadratic band touching at the $\Gamma$ point due to interference effects and real-space connectivity. At the same time, the high degree of symmetry in the atomic arrangement enforces the presence of Dirac points at the BZ boundary, protected by the inherent rotational and mirror symmetries of the lattice. The intricate interplay between geometric deformation and these lattice symmetries fundamentally shapes the electronic band structure of the Kagome lattice, as illustrated in Figure \ref{Kagome_structure}b. This electronic structure can be effectively described by the following periodically modulated tight-binding Hamiltonian:
\begin{equation}
\label{kagome_Hamiltonian}
\mathcal{H}(\bm{k}) = 2t_1\begin{bmatrix}
0 & \cos(\bm{k}\cdot\bm{r}_3) & \cos(\bm{k}\cdot\bm{r}_2)\\
\cos(\bm{k}\cdot\bm{r}_3) & 0 & \cos(\bm{k}\cdot\bm{r}_1) \\
\cos(\bm{k}\cdot\bm{r}_2) & \cos(\bm{k}\cdot\bm{r}_1) & 0
\end{bmatrix},
\end{equation}
where $t_1$ is the hopping energy, three hopping distances are $\bm{r}_1 = \frac{a}{2}(1,0)$, $\bm{r}_2 = \frac{a}{2}\left(-\frac{1}{2}, \frac{\sqrt{3}}{2}\right)$, $\bm{r}_3 = \frac{a}{2}\left(\frac{1}{2}, \frac{\sqrt{3}}{2}\right)$ and $a$ is the lattice constant. This tight-binding model accurately captures the typical band structure of the Kagome lattice, featuring a nearly flat band often located near the Fermi level and a quadratic band touching between this flat band and the upper dispersive band at the $\Gamma$ point.
A constant electric field is applied along the direction of $\bm{b}_1 = \frac{2\pi}{a}\left(1,\frac{\sqrt{3}}{3}\right)$ to simplify the band path of atoms in the vicinity of a quadratic band touching point. And thus, the motion path for carriers in momentum space is defined by $\bm{k} = \alpha (\bm{b}_2 - \bm{b}_1) + \delta (\bm{b}_1 + \bm{b}_2),\ -0.5 \leqslant \alpha \leqslant 0.5$ where $\bm{b}_1 = \frac{2\pi}{a}\left(1,\frac{\sqrt{3}}{3}\right)$, $\bm{b}_2 = \frac{2\pi}{a}\left(0,\frac{2\sqrt{3}}{3}\right)$ and $\delta$ is a small quantity to vary the path around the one exactly crossing $\Gamma$ point. The Hamiltonian along this path is
\begin{equation}
\label{path}
\mathcal{H_{\text{path}}} = 2t_1\begin{bmatrix}
0 & \cos(2\pi\delta) & \cos[\pi(\alpha+\delta)] \\
\cos(2\pi\delta) & 0 & \cos[\pi(\alpha-\delta)] \\
\cos[\pi(\alpha+\delta)] & \cos[\pi(\alpha-\delta)] & 0
\end{bmatrix}.
\end{equation}
The movement path of carriers in momentum space with respect to different values of $\delta$ is plotted by the solid black lines in Figure \ref{fig:Kagome_band}a-c. As the magnitude of $\delta$ increases, the closest approach of the carrier trajectory to the quadratic band touching point at $\Gamma$ moves further away, which corresponds to an increase in the energy difference between the electron-like and hole-like states at the touching point. Furthermore, the opening of a band gap between the two relevant energy bands near $\alpha =0 $ indicates a shift of the exact quadratic band touching point in the Brillouin zone. This opened band gap at the touching point suggests that a Kagome material, similar to the graphene system, inherently facilitates nonadiabatic transitions due to strong tunneling between the electron and hole branches when a small energy difference exists between the two bands at the crossing point.

By linearly expanding the Hamiltonian to the first order around the touching point in the Brillouin zone, near $\alpha = 0$ and $\delta = 0$, and projecting the expanded Hamiltonian onto the subspace spanned by the two eigenvectors of the involved energy levels, the time-dependent Hamiltonian around the quadratic band touching is expressed as follows:
\begin{equation}
\label{expansion}
    \mathcal{H_{\text{expand}}} = 2t_1
    \begin{bmatrix}
        \frac{2}{3}\pi^2\alpha^2-1 & \frac{2\pi^2\alpha\delta}{\sqrt{3}} \\
        \frac{2\pi^2\alpha\delta}{\sqrt{3}} & 2\pi^2\delta^2 - 1
    \end{bmatrix}
\end{equation}
To simplify the notation, we define $\epsilon = \frac{4\pi^2t_1}{3}$, $\tilde{k} = \alpha$, $g = \frac{4\pi^2t_1}{\sqrt{3}}\delta$, ignoring the constant energy offset. Taking the applied electric field into consideration, we have $\hbar\frac{dk}{dt}=e|E|$, where $k=\frac{4\pi}{\sqrt{3}a}\tilde{k}$ in this discussion. This gives $\tilde{k}$ a linear time dependence on the driven field as $\frac{\sqrt{3}ae|E|}{4\pi\hbar}t$. By defining $\beta = \frac{\sqrt{3}ae|E|}{4\pi\hbar}$. Now the time-dependent Hamiltonian is
\begin{equation}
\label{Kagome_expansion}
    \mathcal{{H_{\text{expand}}}} = 
    \begin{bmatrix}
        \epsilon ({\beta t})^2 & g{\beta t} \\
        g{\beta t} & g^2/\epsilon
    \end{bmatrix}.
\end{equation}
The band structure calculated based on this expanded Hamiltonian (Equation \ref{Kagome_expansion}), plotted as red dashed lines in Figure \ref{fig:Kagome_band}a-c, shows excellent agreement with the band structure obtained from the full Hamiltonian (Equation \ref{path}) in the vicinity of the quadratic band touching point for all selected values of $\delta$. This close overlap confirms that the expanded Hamiltonian $H_{\text{expand}}$ provides a highly accurate approximation for the electronic structure and carrier dynamics near the band minimum, validating its use for further analysis of the quadratic band touching

The first order expansion of the time-dependent Hamiltonian \ref{Kagome_expansion} near the quadratic band touching in Kagome systems has a formal resemblance to the standard Zener tunneling Hamiltonian in linearly crossing band systems. However, in this case, the diagonal terms are quadratic with time $t$, while the off-diagonal terms are linear with $t$. Although the LZT are expected to occur near the band touching point, the transition rate and the associated phase shift may exhibit qualitatively different behaviors compared to the conventional linear dispersion scenario.

\begin{figure}[htbp]
\centering
\includegraphics[width=1.0\linewidth]{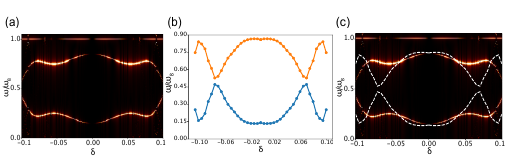}
\caption{Bloch oscillation frequencies as a function of $\delta$ in the Kagome system: (a) extracted from the coherent transport framework, (b) based on the linearly expanded tight-binding model, and (c) comparison between numerical and theoretical results.}
\label{fig:BO_freq_kagome}
\end{figure}

To gain insight into this higher-order band touching, we numerically compute the Bloch oscillation frequency in the Kagome lattice using the coherent transport framework, as presented in Figure \ref{fig:BO_freq_kagome}a. When the parameter $\delta$ approaches zero, the energy gap between the two interacting branches at the quadratic touching point vanishes, resulting in purely adiabatic transitions at the $\Gamma$ point. The periodicity of wavevector propagation within the Brillouin zone, combined with its linear dependence on the applied field, gives rise to a single, regular Bloch oscillation frequency. However, when a finite energy gap opens at the quadratic band touching point, transitions between the two energy branches become possible, leading to a quantum superposition of electron and hole states accompanied by a nontrivial phase accumulation. This interband mixing results in the emergence of two distinct irregular Bloch oscillation frequencies, symmetrically distributed around half of the fundamental Bloch oscillation frequency.
 
Based on the expanded form of the time-dependent Hamiltonian at the quadratic contact point in Equation \ref{Kagome_expansion}, a theoretical framework can be established for understanding the Bloch oscillation frequency in the Kagome system. Following an approach analogous to standard LZT problems, a scattering matrix describing the transitions across the quadratic touching point, accounting for the interaction between the two branches, can be formulated as:
\begin{equation}
\label{scarrtering_matrix}
    \mathcal{S} = 
    \begin{bmatrix}
        {S_{11}}e^{-i{\theta_{11}}} & {S_{12}}e^{-i{\theta_{12}}} \\
        {S_{21}}e^{-i{\theta_{21}}} & {S_{22}}e^{-i{\theta_{22}}}
    \end{bmatrix}
\end{equation}
where $S_{ij}$ represents the probability amplitude for a transition from the j-th energy level to the i-th energy level, and $\theta_{ij}$ denotes the corresponding phase accumulation during the tunneling process. This scattering matrix fully characterizes the key aspects of nonadiabatic transitions at the quadratic touching point.  By dynamically tracking the time evolution of an initial pure state as it traverses the quadratic touching point, evolving into a superposition state, all relevant parameters of the scattering matrix, including the LZT rate and the associated phase accumulation for each carrier trajectory, can be numerically extracted. This approach enables a numerical evaluation of the LZT rate and the associated phase accumulation at the crossing point for each carrier along its prescribed trajectory.

To describe the complete evolution of a state as it traverses the Brillouin zone from $t \rightarrow -\infty$ to $t \rightarrow +\infty$, we consider the total scattering matrix $S_{\text{tot}}$, which incorporates the phase accumulation of the state before and after the tunneling region. These are represented by the time evolution matrices $P$ and $P'$, respectively, such that $S_{\text{tot}} = P' S P$ \cite{sakurai1994modern}. When the flat band quadratically touches the upper energy level without any bandgap, the tunneling probability is calculated to be exactly zero. As a result, the calculated tunneling probability is exactly zero. Consequently, no tunneling-induced transitions occur, and only the fundamental Bloch oscillation frequency is observed. As the effective energy difference between the two interacting bands at the quadratic touching point increases, the Landau-Zener tunneling probability between the electron and hole branches decreases exponentially, following the characteristic suppression due to an increasing energy gap. Diagonalizing the scattering matrix $S_{tot}$ reveals the phase shift between two interacting branches, which indicates the presence of additional oscillation modes arising from the tunneling effect. This tunneling-induced alteration of the periodicity in the carrier motion results in the emergence of two additional Bloch oscillation frequencies, as illustrated in Figure \ref{fig:BO_freq_kagome}b. These frequencies follow a wave-like dependence on the coupling parameter $\delta$, strongly correlated with LZT tunneling dynamics, leading to the appearance of irregular Bloch oscillation frequencies that are close to half the conventional Bloch frequency.

When the coupling parameter between two energy branches, $\delta$, exceeds a threshold of approximately 0.05, the off-diagonal terms of the Hamiltonian become comparable to the diagonal terms. Before this threshold, the diagonalized eigenstates for the expanded Hamiltonian ~\ref{expansion}, given by $\bm{u}_1 = (\frac{-g}{\sqrt{g^2 + (t \beta \epsilon)^2}}, \frac{t \beta \epsilon}{\sqrt{g^2 + (t \beta \epsilon)^2}})$ and $\bm{u}_2 = (\frac{t \beta \epsilon}{\sqrt{g^2 + (t \beta \epsilon)^2}},\frac{g}{\sqrt{g^2 + (t \beta \epsilon)^2}})$, can be approximated as $\begin{bmatrix} 0 \\ 1
\end{bmatrix}$ and $\begin{bmatrix} 1 \\ 0
\end{bmatrix}$ corresponding to the two separate branches at the edges of the Brillouin zone for infinitesimal $\delta$ values. This simplification reflects the behavior of the system when $\delta$ is small, leading to a clear separation between the two branches. However, when $\sqrt{3}$$\delta$ becomes non-negligible, the effective energy difference $2g$ between two branches becomes comparable to energy scale $\epsilon$, indicating strong hybridization between the two branches. In this case, the assumption that the initial state of a carrier entering the Brillouin zone edges can be represented by a single eigenstate of a decoupled branch breaks down. Instead, due to the strong interband coupling, the initial state becomes a coherent superposition of the two branch eigenstates, even before significant evolution under the applied field.  In other words, the assumption that carrier motion across the Brillouin zone can be treated as a simple scattering process from $t=-\infty$ to $t=+\infty$ no longer holds, as the off-diagonal terms significantly influence the dynamics of the system from the outset, including the initial state preparation. Under these conditions, similar to limitations occurred in linear band crossing case with strong coupling, theoretical predictions begin to deviate from the numerical results obtained through coherent transport approaches. This discrepancy is evident in Figure \ref{fig:BO_freq_kagome}c, where the breakdown of the idealized scattering picture leads to a divergence between the analytical and numerical outcomes. The strong initial mixing of states and subsequent coherent evolution within the avoided crossing region are key dynamic features not captured by a pure scattering picture originating from a single initial band.

\section{Lieb system}
\begin{figure}[htbp] 
\centering 
\includegraphics[width=0.6\linewidth]{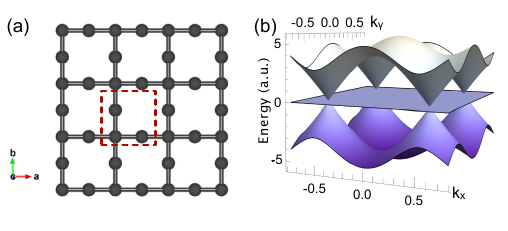} 
\caption{(a) Geometric structure of the Lieb lattice, with the red dashed line marking the unit cell, and (b) its band diagram.} 
\label{Lieb_structure} 
\end{figure}

When the Kagome lattice is subjected to a homogeneous strain of 0.423 along the armchair direction, its hexagonal atomic arrangement undergoes a significant distortion, transforming into the square lattice geometry characteristic of the Lieb lattice. The resulting atomic configuration is illustrated in Figure~\ref{Lieb_structure}a.  This structural transformation disrupts the quadratic touching point between the flat band and the Dirac points. Specifically, the quadratic touching point splits into two distinct Dirac nodes, represented by ($\alpha = \pm\frac{\sqrt{3|\eta|}}{\pi}, \delta = 0$) \cite{jiang_topological_2019}. Furthermore, the flat band is lifted upwards and intersects  the center of these newly formed Dirac points, as clearly illustrated in the electronic band diagram of the Lieb lattice in Figure \ref{Lieb_structure}b.

\begin{figure}[htbp]
\centering
\includegraphics[width=1.0\linewidth]{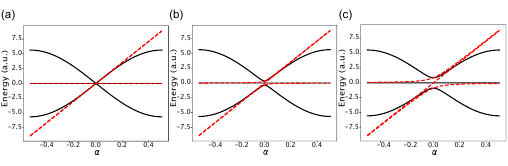}
\caption{ (a-c) Band diagram of the Lieb lattice calculated from Equation~\ref{Lieb_path} (black lines) and Equation~\ref{expansion_lieb_simply} (red dashed lines) when $\delta$ = 0, 0.02 and 0.05, respectively.}
\label{Lieb_band}
\end{figure}

This Lieb lattice system can be effectively modeled using the following periodically modulated tight-binding Hamiltonian:
\begin{equation}
\label{lieb_Hamiltonian}
\mathcal{H} = 2t_1\begin{bmatrix}
0 & \cos(\bm{k}\cdot\bm{r}_2) & \cos(\bm{k}\cdot\bm{r}_1)\\
\cos(\bm{k}\cdot\bm{r}_2) & 0 & 0 \\
\cos(\bm{k}\cdot\bm{r}_1) & 0 & 0
\end{bmatrix},
\end{equation}
where $t_1$ is the hopping energy between nearest-neighbor sites, the relevant hopping distances are $\bm{r}_1 = \frac{a}{2}(0,1)$, $\bm{r}_2 = \frac{a}{2}(1,0)$, and $a$ is the lattice constant. This tight-binding model describes the typical three-band electronic structure of the Lieb lattice, and a constant electric field is applied along the direction of $\bm{b}_2 -\bm{b}_1$ to simplify the band path of the atoms in the vicinity of a band crossing at the $\textbf{M}$ point. And thus, the motion path for carriers in momentum space is defined by $\bm{k} = \alpha (\bm{b}_2-\bm{b}_1) + \delta(\bm{b}_1+\bm{b}_2),\ -0.5 \leqslant \alpha \leqslant 0.5$, where $\bm{b}_1 = \frac{2\pi}{a}\left(1,0\right)$, $\bm{b}_2 = \frac{2\pi}{a}\left(0,1\right)$ and $\delta$ is a small quantity to vary the path around the one passing the band crossing point. The Hamiltonian along this path is
\begin{equation}
\label{Lieb_path}
\mathcal{H_{\text{path}}} = 2t_1\begin{bmatrix}
0 & \cos(\pi(\alpha-\delta) & \cos[\pi(\alpha+\delta)] \\
\cos(\pi(\alpha-\delta) & 0 & 0 \\
\cos[\pi(\alpha+\delta)] & 0 & 0
\end{bmatrix}.
\end{equation}
The path of movement of carriers in momentum space for different values of $\delta$ is shown by the solid black lines in Figure \ref{Lieb_band}a-c. As the magnitude of $\delta$ increases, the closest approach of the carrier trajectory to the Dirac points moves further away, directly reflecting the increasing energy differences between the electron and hole states at these touching points. Furthermore, the opening of a band gap between the intersecting bands lifts the degeneracy of the Dirac points. This opened band gap at the touching point indicates that, like the graphene system, a Lieb lattice inherently supports nonadiabatic transitions. These transitions arise due to strong tunneling between the electron and hole branches when a small energy difference exists between the two bands at the crossing point.

\begin{figure}[htbp]
\centering
\includegraphics[width=1.0\linewidth]{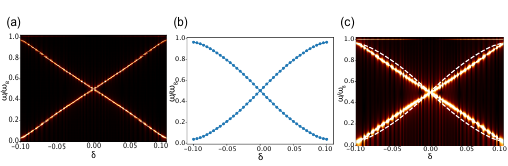}
\caption{Bloch oscillation frequency in the Lieb lattice as a function of $\delta$: (a) calculated from the coherent transport framework, (b) based on the linearly expanded tight-binding model, and (c) comparison between numerical and theoretical results.}
\label{Lieb_theoretical_coherent}
\end{figure}

To analytically investigate the carrier dynamics near the pair of Dirac points,  we perform a linear expansion of the Hamiltonian in Equation \ref{Lieb_path} to the first order near $\alpha = \frac{1}{2}$ and $\delta = 0$. This yields the following effective time-dependent Hamiltonian near the pair of Dirac points:
\begin{equation}
\label{expansion_lieb}
    \mathcal{H_{\text{expand}}} = 2\pi t_1
    \begin{bmatrix}
        \sqrt{2}(\alpha-\frac{1}{2}) & 0 & \delta \\ 0 & -\sqrt{2}(\alpha-\frac{1}{2}) & \delta \\ \delta & \delta & 0
    \end{bmatrix}.
\end{equation}
To have a simplified notation of the Hamiltonian in Equation~\ref{expansion_lieb}, notations defined as $\epsilon_0 = 2\sqrt{2}\pi t_1$, $\tilde{k} = \alpha - \frac{1}{2}$, $g =2\pi t_1\delta$ and the overall offset of the bands is ignored. Taking the applied electric field into consideration, we have $\hbar\frac{dk}{dt}=e|E|$, where $k=\frac{2\sqrt{2}\pi}{a}\tilde{k}$ in this discussion. This gives $\tilde{k}$ a linear time dependence on the driven field as $\frac{ae|E|}{2\sqrt{2}\pi\hbar}t$. By defining $\beta = \frac{ae|E|}{2\sqrt{2}\pi\hbar}$ and $\epsilon = \epsilon_0\beta$, we have
\begin{equation}
\label{expansion_lieb_simply}
\mathcal{{H_{\text{expand}}}} = 
    \begin{bmatrix}
        \epsilon t & 0 & g \\
        0 & -\epsilon t & g\\
        g & g & 0
    \end{bmatrix}.
\end{equation}
The band diagrams of the Lieb lattice, derived from the tight-binding model in Equation~\ref{Lieb_path} and the expanded Lieb Hamiltonian in The electronic band diagrams of the Lieb lattice, calculated from the full tight-binding model (Equation \ref{Lieb_path}, black solid lines) and the expanded Hamiltonian (Equation \ref{expansion_lieb_simply}, red dashed lines), are presented in Figure \ref{Lieb_band} for various values of $\delta$. As shown, increasing $\delta$ results in a gradual increase in the energy separation between the three bands. In all cases, the expanded Hamiltonian shows excellent agreement with the original Lieb lattice Hamiltonian, confirming its accuracy in capturing the essential electronic structure relevant for analyzing Bloch oscillations near the band crossings.
Following the same numerical approach used for the Kagome system, Bloch oscillations in the Lieb lattice are evaluated by implementing the coherent transport framework based on its tight-binding model. Unlike the Kagome lattice, where Dirac points and a flat band cross at multiple points in the Brillouin zone, the Lieb lattice features three energy bands where the middle flat band intersects the two dispersive bands at the $M$ point, forming a triply degenerate Dirac-like crossing at zero energy. However, a detailed examination of the expanded Hamiltonian in Equation~\ref{expansion_lieb_simply} reveals crucially that the two dispersive bands interact independently with the flat band, with no direct interband coupling between the two dispersive bands themselves. Consequently, the system can be effectively described as two pairs of linear crossings, each comprising a dispersive band and a dispersionless flat band, occurring effectively at the $\Gamma$ point in the Brillouin zone. When $\delta$ is zero, corresponding to the exact band touching, LZT does not directly influence the evolution of pure states at the crossing points. As a result, the crossing at the $M$ point doubles the periodicity, resulting in a Bloch oscillation frequency that is halved compared to the fundamental frequency expected from a single isolated band. For nonzero values of $\delta$, the energy gap between these pairs of interacting bands increases symmetrically. Two independent LZT events occur directly between each of the linear dispersion bands and the flat band. Additionally, the LZT between the two dispersive bands arises from second-order processes involving successive hopping events, first tunneling from a dispersive band to the flat band, followed by tunneling from the flat band to the other dispersive band. This unique coupling scheme results in an irregular Bloch oscillation frequency that exhibits a linear dependence on the energy difference between the crossing branches, as shown in Figure~\ref{Lieb_theoretical_coherent}a.

The time-dependent Hamiltonian expansion in the Lieb lattice deviates from the standard Zener tunneling in systems with linear band crossings due to the increased dimensionality and the presence of a flat band, which modify the tunneling dynamics and the overall energy dispersion. Consequently, the differential equations governing the time evolution of the state wavefunctions involve special functions that are mathematically complex, with further details provided in the supporting information, making their analytical presentation and direct interpretation challenging. Adopting a similar approach to the Kagome lattice study, key parameters for constructing the three-dimensional scattering matrix, such as scattering probabilities and accumulated phase, are numerically determined by tracking the time evolution of initial states governed by the expanded Lieb Hamiltonian. The Bloch oscillation frequency, extracted from scattering phases, is represented by blue dots in Figure~\ref{Lieb_theoretical_coherent}b. The comparison between the irregular Bloch oscillations predicted by the coherent framework and those obtained from the scattering model are illustrated in Figure~\ref{Lieb_theoretical_coherent}c. Both approaches consistently demonstrate a linear dependence of the irregular Bloch oscillation frequency on the energy gap associated with the LZT process. For values of $\delta$ greater than 0.1, the energy gap between the interacting branches becomes sufficiently large to effectively suppress interband interactions, leading to the dominance of fundamental Bloch oscillations beyond this threshold.

\section{Strain induced transformation between Lieb and Kagome system}
Strain engineering in 2D materials has attracted significant interest as its power for modifying mechanical and electronic properties through geometric deformation \cite{dai_strain_2019}. A compelling example of this is the structural transformation from the Kagome lattice to the Lieb lattice under applied external strain, where the strain parameters act as a continuous tuning parameter. This transition involves a change from the corner-sharing triangular arrangement of the Kagome lattice to the square-based Lieb lattice featuring centered atomic sites, leading to substantial modifications in the electronic band structure. Despite these alterations, key topological invariants, such as Chern numbers, topological phases, and edge states, remain preserved, highlighting the resilience of the underlying topology characteristics of the system. Within this dynamically evolving framework, Bloch oscillations, a fundamental quantum phenomenon arising from particle motion in a periodic potential, take on a new level of complexity. As the lattice undergoes transformation, these oscillations may experience dynamic transitions, reflecting the continuous evolution from the Kagome to the Lieb structure.

To numerically explore the impact of strain on the irregular Bloch oscillation frequency, an external strain $\eta$ is applied along the $\bm{b}_2 - \bm{b}_1$ direction, which is aligned with the direction of the applied electric field. The strain is integrated into the tight-binding model of the Kagome lattice by modifying the lattice vectors and the atomic positions within the unit cell according to the following relations:
\begin{equation}
\begin{aligned}
    &\bm{r}_1=\frac{a}{2}\left(1+\frac{3\eta}{4},-\frac{\sqrt{3}\eta}{4}\right),\ 
    \bm{r}_2=\frac{a}{2}\left(-\frac{1}{2}-\frac{3\eta}{4},\frac{\sqrt{3}}{2}+\frac{\sqrt{3}\eta}{4}\right),\ 
    \bm{r}_3=\frac{a}{2}\left(\frac{1}{2},\frac{\sqrt{3}}{2}\right)
    \\&\bm{a}_1 = a\left(1+\frac{3\eta}{4},-\frac{\sqrt{3}\eta}{4}\right),\ 
    \bm{a}_2 = a\left(-\frac{1}{2}-\frac{3\eta}{4},\frac{\sqrt{3}}{2}+\frac{\sqrt{3}\eta}{4}\right)
    \\&\bm{b}_1 = \frac{\pi}{a(1+\eta)}\left(2+\eta,\frac{2+3\eta}{\sqrt{3}}\right),\ 
    \bm{b}_2 = \frac{\pi}{a(1+\eta)}\left(\eta,\frac{4+3\eta}{\sqrt{3}}\right).
\end{aligned}
\end{equation}
These result in changes in the lattice structure, through changes in the bond length due to strain. The updated hopping energy is related to the original value by the expression $t'=t_1\left(\frac{r_0}{r}\right)^2$, where $r_0$ is the bond length of the bond in the unstrained structure, and $r$ is the bond length in the strained structure. The particle motion paths remain along the direction of the applied field and are defined as \begin{equation} 
\bm{k} = \alpha (\bm{b}_2-\bm{b}_1) + \delta(\bm{b}_1+\bm{b}_2),\ -0.5 \leqslant \alpha \leqslant 0.5. 
\end{equation}
The tight-binding Hamiltonian for the strained Kagome lattice, evaluated along the defined carrier motion path in momentum space, is given by:
\begin{equation}
\label{Hamiltonian_strain_1}
   \mathcal{H} = 2t'\begin{bmatrix}
0 & \cos(2\pi\delta) & \frac{4\cos[\pi(\alpha+\delta)]}{4+3\eta(2+\eta)} \\
\cos(2\pi\delta) & 0 &\frac{4\cos[\pi(\alpha-\delta)]}{4+3\eta(2+\eta)}\\ \frac{4\cos[\pi(\alpha+\delta)]}{4+3\eta(2+\eta)} & \frac{4\cos[\pi(\alpha-\delta)]}{4+3\eta(2+\eta)} & 0
\end{bmatrix}.
\end{equation}
Linearly expanding this Hamitonian at the quadratic touching point, $\Gamma$, and then projecting into the two dimensional subspace of interest, the resulting LZT Hamiltonian becomes:
\begin{equation}
\label{expansion_strain_1}
    \mathcal{H}_{\text{expand}}
    = 2t'
    \begin{bmatrix}
        \frac{2}{3}\pi^2\alpha^2+2\eta - 1 & \frac{2\pi^2\alpha\delta}{\sqrt{3}} \\
        \frac{2\pi^2\alpha\delta}{\sqrt{3}} & 2\pi^2\delta^2 - 1
    \end{bmatrix}.
\end{equation}
A crucial observation from this expanded Hamiltonian is that, while strain induces significant modifications to the diagonal terms, specifically introducing an additional energy shift of 2$\eta$, the off-diagonal terms, which are solely responsible for Landau-Zener tunneling between the two energy branches, remain unchanged compared to the unstrained case. This suggests that the fundamental Landau-Zener transition probability and the associated tunneling dynamics are intrinsically robust against the applied biaxial strain, even though the overall band structure is significantly altered. 

\begin{figure}[htbp]
\centering
\includegraphics[width=0.6\linewidth]{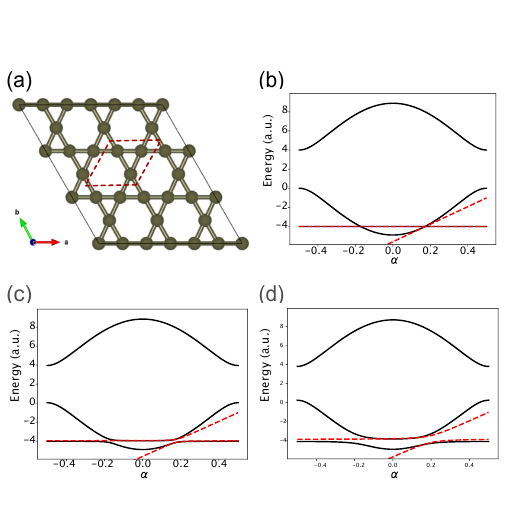}
\caption{(a) The top view of \( 3 \times 3 \times 1 \) Kagome super lattice structure under 0.1 compressive strain along the diagonal direction, and (b-d) its corresponding band structure calculated from Equation~\ref{expansion_strain_1} (black lines) and Equation~\ref{expansion_Dirac_nodes} (red dashed lines) for $\delta$ = 0, 0.1 and 0.2, respectively.}
\label{fig:Band_structure_strain}
\end{figure}

When the strain is compressive ($\eta$$<$0), the flat band gradually shifts away from the Fermi level as the strain increases. For example, in the Kagome lattice under 0.1 compression, the corresponding band structure obtained from Eq.~\ref{expansion_strain_1} is shown by the black lines in Fig.~\ref{fig:Band_structure_strain}b. In contrast to the unstrained Kagome system, the quadratic band touching is disrupted, and the flat band is clearly lifted above its original position. The band structure now exhibits two adjacent linear crossing points, symmetric about the $\Gamma$ point, replacing the single quadratic touching. Expanding the strained Hamiltonian around these new crossing points, the resulting Landau-Zener transition Hamiltonian can be expressed as follows:
\begin{equation}
\label{expansion_Dirac_nodes}
    \mathcal{H}_{LZ} = 2t_1
    \begin{bmatrix}
        \frac{4\pi\sqrt{|\eta|}}{\sqrt{3}}\left(\alpha\mp\frac{\sqrt{3|\eta|}}{\pi}\right)- 1 & 2\pi\delta\sqrt{|\eta|} \\
        2\pi\delta\sqrt{|\eta|} &  2\pi^2\delta^2- 1
    \end{bmatrix}.
\end{equation}
This represents a prototype Hamiltonian for the Landau-Zener transition in the context of linear crossings, when the off-diagonal term is assigned nonzero values, a bandgap opens between the two branches at the crossing points. As shown by the solid lines in Figure \ref {fig:Band_structure_strain}b-c, with the off-diagonal term assigned nonzero values, a bandgap opens between the two branches at the crossing points. The expanded Hamiltonian, illustrated by the red dashed lines, demonstrates that this linear expansion closely follows the original band structure, regardless of whether a bandgap forms between the electron and hole branches.

\begin{figure}[htbp]
\centering
\includegraphics[width=1.0\linewidth]
{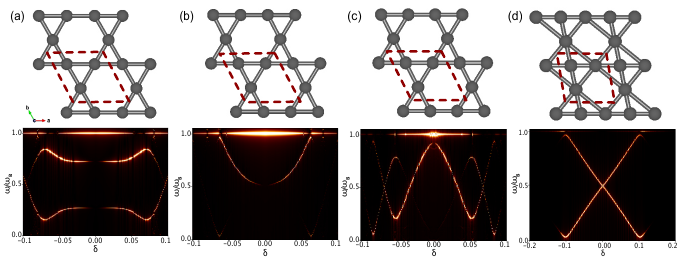}
\caption{Schematic geometry configurations of the compressed Kagome lattice (top) and calculated Bloch oscillation frequency via coherent transport framework as a function of $\delta$ (bottom), with applied compressive strains of (a) $\eta=-0.01$, (b) $\eta=-0.05$, (c) $\eta=-0.10$, and (d) $\eta=-0.29$.}
\label{fig:BO_freq_strain_2}
\end{figure}

The effect of strain on Bloch oscillation frequencies in the Kagome lattice is dynamically evaluated for varying strain levels using the coherent transport framework, and the result is presented in Figure \ref{fig:BO_freq_strain_2}. The numerical simulations reveal the emergence of two irregular Bloch oscillation frequencies, originating from Landau-Zener tunneling between the electron and hole branches, across different applied strain values. For compressive strains with a magnitude less than approximately 0.03, the pattern of the Bloch oscillation frequencies as a function of the tunneling parameter $\delta$ closely resembles that observed in the unstrained Kagome system, characterized by the waving pattern around half the fundamental frequency. As the strain increases, the BO frequency for the unavoided crossing tends toward half of the fundamental value. This happens because strain deforms the quadratic band-touching point and changes the band dispersion, though at small strain it is not strong enough to fully remove the crossing. Once the compressive strain exceeds about $\eta=-0.05$, the BO frequency pattern begins to deviate noticeably from that of the unstrained Kagome lattice. At this stage, the flat band is lifted from its original position, and the quadratic band touching splits into a pair of Dirac-like linear crossings. As the strain approaches $\eta=-0.29$, the lattice angle shifts from the hexagonal value toward the rectangular geometry of the Lieb lattice. In this regime, the stronger lifting of the bands produces a doubled real-space periodicity of the BOs, giving a frequency close to half the fundamental value, consistent with the behavior expected for the Lieb lattice. Consistent with the first-order expansion of the strained Kagome Hamiltonian at the band touching points, which showed that strain does not introduce additional terms affecting the LZT probability, the evolution of the Bloch oscillation frequency pattern from the Kagome to the Lieb regime can be interpreted as a gradual compression and modification of the initial Kagome oscillation pattern into that characteristic of the Lieb lattice as the applied strain increases.

\section{Conclusions and Future Directions}
In this study, we investigated the influence of Landau-Zener transitions (LZTs) on Bloch oscillation (BO) frequencies in three-level energy systems, specifically the Kagome and Lieb lattices, under a constant electric field. Our investigation employed two complementary approaches, a numerical coherent transport framework based on the tight-binding model and a theoretical scattering matrix derived from the scattering Hamiltonian. As a benchmark, we first analyzed the graphene system using these two independent methodologies. Both yielded consistent results, identifying two irregular Bloch oscillation frequencies that displayed an interwoven pattern around half the fundamental frequency, a direct consequence of the LZT effect. This successful validation allowed us to extend our investigation to the more complex Kagome and Lieb lattices.

Our analysis, based on the scattering Hamiltonian, reveals a key similarity between these flat band systems and graphene, which is that the off-diagonal terms in the effective Hamiltonian are the primary determinants of LZT occurrence in both the Kagome and Lieb lattices. These terms are directly proportional to the energy difference between the interacting bands at the avoided crossing. Notably, this fundamental contribution to the LZT probability remains invariant under strain during the geometric transformation between the two lattice types. Furthermore, scattering matrix analysis indicates distinct LZT pathways in each lattice. In the Kagome lattice, LZT predominantly involve the dispersive bands near the quadratic band touching with the flat band. Conversely, in the Lieb lattice, the two dispersive bands, characterized by linear dispersion, interact independently with the flat band to produce LZTs, with no direct tunneling observed between these two dispersive bands at the energy scales relevant to these transitions. In the adiabatic condition for the unavoided band touching, we observed a fundamental Bloch oscillation frequency in the Kagome lattice and a half-fundamental frequency in the Lieb lattice, which is a direct reflection from their differing electronic band structures. When a small energy difference was introduced at the band touching point, creating an avoided crossing and enabling non-adiabatic transitions, both lattices exhibited irregular BO frequencies. These irregular frequencies, calculated by both our coherent transport framework and scattering matrix method, showed a waving pattern as a function of coupling parameters in the Kagome lattice and a linear dependence on the energy difference in the Lieb lattice. Although the intrinsic nature of the Landau-Zener transition, governed by the off-diagonal coupling in the effective Hamiltonian, remains robust against strain, the specific patterns of the resulting irregular BOs clearly depended on the applied strain. Our dynamic simulations, using the coherent transport framework, further demonstrated that a strain-induced geometric transformation from the Kagome to the Lieb lattice led to a compression of both the lattice geometry and the characteristic patterns of the irregular BO frequencies.

Our numerical evaluations and scattering matrix analysis of irregular Bloch oscillations offer deeper insights into non-adiabatic quantum transport in three-level energy systems, especially within topological and flat-band materials. These findings have important implications for understanding and potentially controlling optical responses in tunable photonic devices, providing insight into the role of irregular oscillation patterns in manipulating light-matter interactions. Moreover, our study highlights the resilience of the fundamental LZT dynamics, governed by the off-diagonal terms in the effective Landau-Zener Hamiltonian, against strain-induced geometric transformations between the Lieb and Kagome lattices. This persistent LZT effect, largely unchanged by strain, suggests a potential mechanism for rapid modulation of electronic states in these materials. When combined with external fields, it could enable controlled transitions between distinct electronic phases, including those with different topological properties and band structures. This holds relevance for developing topological qubits, where precise control over electronic states and phase coherence is essential for quantum computing applications.

\begin{acknowledgments}
This work was primarily supported by grant DE-SC0023432 funded by the U.S. Department of Energy, Office of Science. This research used resources of the National Energy Research Scientific Computing Center, a DOE Office of Science User Facility supported by the Office of Science of the U.S. Department of Energy under Contract No.~DE-AC02-05CH11231, using NERSC awards BES-ERCAP0033206, BES-ERCAP0025205, BES-ERCAP0025168, and BES-ERCAP0028072. Additionally, ASB acknowledges startup support from the Samueli School Of Engineering at UCLA. CJGC acknowledges support from the AFOSR through grant FA9550-18-1-0095. The authors acknowledge the use of the GPT-5 (OpenAI) model to polish the language and edit grammatical errors in some sections of this manuscript. The authors subsequently inspected, validated and edited the text generated by the AI model, before incorporation. \end{acknowledgments}
\bibliography{BO_reference}

\begin{thebibliography}{46}%
\makeatletter
\providecommand \@ifxundefined [1]{%
 \@ifx{#1\undefined}
}%
\providecommand \@ifnum [1]{%
 \ifnum #1\expandafter \@firstoftwo
 \else \expandafter \@secondoftwo
 \fi
}%
\providecommand \@ifx [1]{%
 \ifx #1\expandafter \@firstoftwo
 \else \expandafter \@secondoftwo
 \fi
}%
\providecommand \natexlab [1]{#1}%
\providecommand \enquote  [1]{``#1''}%
\providecommand \bibnamefont  [1]{#1}%
\providecommand \bibfnamefont [1]{#1}%
\providecommand \citenamefont [1]{#1}%
\providecommand \href@noop [0]{\@secondoftwo}%
\providecommand \href [0]{\begingroup \@sanitize@url \@href}%
\providecommand \@href[1]{\@@startlink{#1}\@@href}%
\providecommand \@@href[1]{\endgroup#1\@@endlink}%
\providecommand \@sanitize@url [0]{\catcode `\\12\catcode `\$12\catcode `\&12\catcode `\#12\catcode `\^12\catcode `\_12\catcode `\%12\relax}%
\providecommand \@@startlink[1]{}%
\providecommand \@@endlink[0]{}%
\providecommand \url  [0]{\begingroup\@sanitize@url \@url }%
\providecommand \@url [1]{\endgroup\@href {#1}{\urlprefix }}%
\providecommand \urlprefix  [0]{URL }%
\providecommand \Eprint [0]{\href }%
\providecommand \doibase [0]{http://dx.doi.org/}%
\providecommand \selectlanguage [0]{\@gobble}%
\providecommand \bibinfo  [0]{\@secondoftwo}%
\providecommand \bibfield  [0]{\@secondoftwo}%
\providecommand \translation [1]{[#1]}%
\providecommand \BibitemOpen [0]{}%
\providecommand \bibitemStop [0]{}%
\providecommand \bibitemNoStop [0]{.\EOS\space}%
\providecommand \EOS [0]{\spacefactor3000\relax}%
\providecommand \BibitemShut  [1]{\csname bibitem#1\endcsname}%
\let\auto@bib@innerbib\@empty
\bibitem [{\citenamefont {Bloch}(1929)}]{bloch_uber_1929}%
  \BibitemOpen
  \bibfield  {author} {\bibinfo {author} {\bibfnamefont {F.}~\bibnamefont {Bloch}},\ }\href {\doibase 10.1007/BF01339455} {\bibfield  {journal} {\bibinfo  {journal} {Zeitschrift f{\"u}r Physik}\ }\textbf {\bibinfo {volume} {52}},\ \bibinfo {pages} {555} (\bibinfo {year} {1929})}\BibitemShut {NoStop}%
\bibitem [{\citenamefont {Zener}\ and\ \citenamefont {Fowler}(1934)}]{zener1934theory}%
  \BibitemOpen
  \bibfield  {author} {\bibinfo {author} {\bibfnamefont {C.}~\bibnamefont {Zener}}\ and\ \bibinfo {author} {\bibfnamefont {R.~H.}\ \bibnamefont {Fowler}},\ }\href {\doibase 10.1098/rspa.1934.0116} {\bibfield  {journal} {\bibinfo  {journal} {Proceedings of the Royal Society of London. Series A, Containing Papers of a Mathematical and Physical Character}\ }\textbf {\bibinfo {volume} {145}},\ \bibinfo {pages} {523} (\bibinfo {year} {1934})}\BibitemShut {NoStop}%
\bibitem [{\citenamefont {Ziman}(1972)}]{ziman_principles_1972}%
  \BibitemOpen
  \bibfield  {author} {\bibinfo {author} {\bibfnamefont {J.~M.}\ \bibnamefont {Ziman}},\ }\href {\doibase 10.1017/CBO9781139644075} {\emph {\bibinfo {title} {Principles of the Theory of Solids}}},\ \bibinfo {edition} {2nd}\ ed.\ (\bibinfo  {publisher} {Cambridge University Press},\ \bibinfo {year} {1972})\BibitemShut {NoStop}%
\bibitem [{\citenamefont {Wannier}(1960)}]{wannier1960effective}%
  \BibitemOpen
  \bibfield  {author} {\bibinfo {author} {\bibfnamefont {G.~H.}\ \bibnamefont {Wannier}},\ }\href {\doibase 10.1103/PhysRev.117.432} {\bibfield  {journal} {\bibinfo  {journal} {Physical Review}\ }\textbf {\bibinfo {volume} {117}},\ \bibinfo {pages} {432} (\bibinfo {year} {1960})}\BibitemShut {NoStop}%
\bibitem [{\citenamefont {Mendez}\ and\ \citenamefont {Bastard}(1993)}]{mendez1993wannier}%
  \BibitemOpen
  \bibfield  {author} {\bibinfo {author} {\bibfnamefont {E.~E.}\ \bibnamefont {Mendez}}\ and\ \bibinfo {author} {\bibfnamefont {G.}~\bibnamefont {Bastard}},\ }\href {\doibase 10.1063/1.881353} {\bibfield  {journal} {\bibinfo  {journal} {Physics Today}\ }\textbf {\bibinfo {volume} {46}},\ \bibinfo {pages} {34} (\bibinfo {year} {1993})}\BibitemShut {NoStop}%
\bibitem [{\citenamefont {Leo}\ \emph {et~al.}(1992)\citenamefont {Leo}, \citenamefont {Bolívar}, \citenamefont {Brüggemann},\ and\ \citenamefont {Schwedler}}]{LEO1992943}%
  \BibitemOpen
  \bibfield  {author} {\bibinfo {author} {\bibfnamefont {K.}~\bibnamefont {Leo}}, \bibinfo {author} {\bibfnamefont {P.~H.}\ \bibnamefont {Bolívar}}, \bibinfo {author} {\bibfnamefont {F.}~\bibnamefont {Brüggemann}}, \ and\ \bibinfo {author} {\bibfnamefont {R.}~\bibnamefont {Schwedler}},\ }\href {\doibase 10.1016/0038-1098(92)90798-E} {\bibfield  {journal} {\bibinfo  {journal} {Solid State Communications}\ }\textbf {\bibinfo {volume} {84}},\ \bibinfo {pages} {943} (\bibinfo {year} {1992})}\BibitemShut {NoStop}%
\bibitem [{\citenamefont {Dahan}\ \emph {et~al.}(1996)\citenamefont {Dahan}, \citenamefont {Peik}, \citenamefont {Reichel}, \citenamefont {Castin},\ and\ \citenamefont {Salomon}}]{PhysRevLett.76.4508}%
  \BibitemOpen
  \bibfield  {author} {\bibinfo {author} {\bibfnamefont {M.~B.}\ \bibnamefont {Dahan}}, \bibinfo {author} {\bibfnamefont {E.}~\bibnamefont {Peik}}, \bibinfo {author} {\bibfnamefont {J.}~\bibnamefont {Reichel}}, \bibinfo {author} {\bibfnamefont {Y.}~\bibnamefont {Castin}}, \ and\ \bibinfo {author} {\bibfnamefont {C.}~\bibnamefont {Salomon}},\ }\href {\doibase 10.1103/PhysRevLett.76.4508} {\bibfield  {journal} {\bibinfo  {journal} {Physical Review Letters}\ }\textbf {\bibinfo {volume} {76}},\ \bibinfo {pages} {4508} (\bibinfo {year} {1996})}\BibitemShut {NoStop}%
\bibitem [{\citenamefont {Haller}\ \emph {et~al.}(2010)\citenamefont {Haller}, \citenamefont {Hart}, \citenamefont {Mark}, \citenamefont {Danzl}, \citenamefont {Reichsöllner},\ and\ \citenamefont {Nägerl}}]{haller2010inducing}%
  \BibitemOpen
  \bibfield  {author} {\bibinfo {author} {\bibfnamefont {E.}~\bibnamefont {Haller}}, \bibinfo {author} {\bibfnamefont {R.}~\bibnamefont {Hart}}, \bibinfo {author} {\bibfnamefont {M.~J.}\ \bibnamefont {Mark}}, \bibinfo {author} {\bibfnamefont {J.~G.}\ \bibnamefont {Danzl}}, \bibinfo {author} {\bibfnamefont {L.}~\bibnamefont {Reichsöllner}}, \ and\ \bibinfo {author} {\bibfnamefont {H.-C.}\ \bibnamefont {Nägerl}},\ }\href {\doibase 10.1103/PhysRevLett.104.200403} {\bibfield  {journal} {\bibinfo  {journal} {Physical Review Letters}\ }\textbf {\bibinfo {volume} {104}},\ \bibinfo {pages} {200403} (\bibinfo {year} {2010})}\BibitemShut {NoStop}%
\bibitem [{\citenamefont {Guo}\ \emph {et~al.}(2021)\citenamefont {Guo}, \citenamefont {Ge}, \citenamefont {Li}, \citenamefont {Wang}, \citenamefont {Zhang}, \citenamefont {Song}, \citenamefont {Xiang}, \citenamefont {Song}, \citenamefont {Jin}, \citenamefont {Lu}, \citenamefont {Xu}, \citenamefont {Zheng},\ and\ \citenamefont {Fan}}]{guo2021observation}%
  \BibitemOpen
  \bibfield  {author} {\bibinfo {author} {\bibfnamefont {X.-Y.}\ \bibnamefont {Guo}}, \bibinfo {author} {\bibfnamefont {Z.-Y.}\ \bibnamefont {Ge}}, \bibinfo {author} {\bibfnamefont {H.}~\bibnamefont {Li}}, \bibinfo {author} {\bibfnamefont {Z.}~\bibnamefont {Wang}}, \bibinfo {author} {\bibfnamefont {Y.-R.}\ \bibnamefont {Zhang}}, \bibinfo {author} {\bibfnamefont {P.}~\bibnamefont {Song}}, \bibinfo {author} {\bibfnamefont {Z.}~\bibnamefont {Xiang}}, \bibinfo {author} {\bibfnamefont {X.}~\bibnamefont {Song}}, \bibinfo {author} {\bibfnamefont {Y.}~\bibnamefont {Jin}}, \bibinfo {author} {\bibfnamefont {L.}~\bibnamefont {Lu}}, \bibinfo {author} {\bibfnamefont {K.}~\bibnamefont {Xu}}, \bibinfo {author} {\bibfnamefont {D.}~\bibnamefont {Zheng}}, \ and\ \bibinfo {author} {\bibfnamefont {H.}~\bibnamefont {Fan}},\ }\href {\doibase 10.1038/s41534-021-00385-3} {\bibfield  {journal} {\bibinfo  {journal} {npj Quantum Information}\ }\textbf {\bibinfo {volume} {7}},\ \bibinfo {pages} {51} (\bibinfo {year} {2021})}\BibitemShut
  {NoStop}%
\bibitem [{\citenamefont {Tans}\ \emph {et~al.}(1997)\citenamefont {Tans}, \citenamefont {Devoret}, \citenamefont {Dai}, \citenamefont {Thess}, \citenamefont {Smalley}, \citenamefont {Geerligs},\ and\ \citenamefont {Dekker}}]{tans1997individual}%
  \BibitemOpen
  \bibfield  {author} {\bibinfo {author} {\bibfnamefont {S.~J.}\ \bibnamefont {Tans}}, \bibinfo {author} {\bibfnamefont {M.~H.}\ \bibnamefont {Devoret}}, \bibinfo {author} {\bibfnamefont {H.}~\bibnamefont {Dai}}, \bibinfo {author} {\bibfnamefont {A.}~\bibnamefont {Thess}}, \bibinfo {author} {\bibfnamefont {R.~E.}\ \bibnamefont {Smalley}}, \bibinfo {author} {\bibfnamefont {L.~J.}\ \bibnamefont {Geerligs}}, \ and\ \bibinfo {author} {\bibfnamefont {C.}~\bibnamefont {Dekker}},\ }\href {\doibase 10.1038/386474a0} {\bibfield  {journal} {\bibinfo  {journal} {Nature}\ }\textbf {\bibinfo {volume} {386}},\ \bibinfo {pages} {474} (\bibinfo {year} {1997})}\BibitemShut {NoStop}%
\bibitem [{\citenamefont {Haug}\ and\ \citenamefont {Koch}(2004)}]{haug2009quantum}%
  \BibitemOpen
  \bibfield  {author} {\bibinfo {author} {\bibfnamefont {H.}~\bibnamefont {Haug}}\ and\ \bibinfo {author} {\bibfnamefont {S.}~\bibnamefont {Koch}},\ }\href {\doibase 10.1063/1.2808410} {\emph {\bibinfo {title} {Quantum Theory of Optical and Electronic Properties of Semiconductors}}},\ Vol.~\bibinfo {volume} {47}\ (\bibinfo  {publisher} {World Scientific},\ \bibinfo {year} {2004})\BibitemShut {NoStop}%
\bibitem [{\citenamefont {Raizen}\ \emph {et~al.}(1997)\citenamefont {Raizen}, \citenamefont {Salomon},\ and\ \citenamefont {Niu}}]{raizen1997new}%
  \BibitemOpen
  \bibfield  {author} {\bibinfo {author} {\bibfnamefont {M.}~\bibnamefont {Raizen}}, \bibinfo {author} {\bibfnamefont {C.}~\bibnamefont {Salomon}}, \ and\ \bibinfo {author} {\bibfnamefont {Q.}~\bibnamefont {Niu}},\ }\href {\doibase 10.1063/1.881845} {\bibfield  {journal} {\bibinfo  {journal} {Physics Today}\ }\textbf {\bibinfo {volume} {50}},\ \bibinfo {pages} {30} (\bibinfo {year} {1997})}\BibitemShut {NoStop}%
\bibitem [{\citenamefont {Zhang}\ and\ \citenamefont {Liu}(2010)}]{zhang2010directed}%
  \BibitemOpen
  \bibfield  {author} {\bibinfo {author} {\bibfnamefont {J.~M.}\ \bibnamefont {Zhang}}\ and\ \bibinfo {author} {\bibfnamefont {W.~M.}\ \bibnamefont {Liu}},\ }\href {\doibase 10.1103/PhysRevA.82.025602} {\bibfield  {journal} {\bibinfo  {journal} {Phys. Rev. A}\ }\textbf {\bibinfo {volume} {82}},\ \bibinfo {pages} {025602} (\bibinfo {year} {2010})}\BibitemShut {NoStop}%
\bibitem [{\citenamefont {Sun}\ \emph {et~al.}(2018)\citenamefont {Sun}, \citenamefont {Leykam}, \citenamefont {Nenni}, \citenamefont {Song}, \citenamefont {Chen}, \citenamefont {Chong},\ and\ \citenamefont {Chen}}]{sun_observation_2018}%
  \BibitemOpen
  \bibfield  {author} {\bibinfo {author} {\bibfnamefont {Y.}~\bibnamefont {Sun}}, \bibinfo {author} {\bibfnamefont {D.}~\bibnamefont {Leykam}}, \bibinfo {author} {\bibfnamefont {S.}~\bibnamefont {Nenni}}, \bibinfo {author} {\bibfnamefont {D.}~\bibnamefont {Song}}, \bibinfo {author} {\bibfnamefont {H.}~\bibnamefont {Chen}}, \bibinfo {author} {\bibfnamefont {Y.}~\bibnamefont {Chong}}, \ and\ \bibinfo {author} {\bibfnamefont {Z.}~\bibnamefont {Chen}},\ }\href {\doibase 10.1103/PhysRevLett.121.033904} {\bibfield  {journal} {\bibinfo  {journal} {Phys. Rev. Lett.}\ }\textbf {\bibinfo {volume} {121}},\ \bibinfo {pages} {033904} (\bibinfo {year} {2018})}\BibitemShut {NoStop}%
\bibitem [{\citenamefont {Dittrich}\ \emph {et~al.}(1998)\citenamefont {Dittrich}, \citenamefont {Hänggi}, \citenamefont {Ingold}, \citenamefont {Kramer}, \citenamefont {Schön},\ and\ \citenamefont {Zwerger}}]{dittrich1998quantum}%
  \BibitemOpen
  \bibfield  {author} {\bibinfo {author} {\bibfnamefont {T.}~\bibnamefont {Dittrich}}, \bibinfo {author} {\bibfnamefont {P.}~\bibnamefont {Hänggi}}, \bibinfo {author} {\bibfnamefont {G.-L.}\ \bibnamefont {Ingold}}, \bibinfo {author} {\bibfnamefont {B.}~\bibnamefont {Kramer}}, \bibinfo {author} {\bibfnamefont {G.}~\bibnamefont {Schön}}, \ and\ \bibinfo {author} {\bibfnamefont {W.}~\bibnamefont {Zwerger}},\ }\href {https://www.researchgate.net/publication/266041870_Quantum_Transport_and_Dissipation} {\emph {\bibinfo {title} {Quantum Transport and Dissipation}}},\ \bibinfo {edition} {1st}\ ed.\ (\bibinfo  {publisher} {Wiley-VCH},\ \bibinfo {year} {1998})\BibitemShut {NoStop}%
\bibitem [{\citenamefont {Lim}\ \emph {et~al.}(2012)\citenamefont {Lim}, \citenamefont {Fuchs},\ and\ \citenamefont {Montambaux}}]{PhysRevLett.108.175303}%
  \BibitemOpen
  \bibfield  {author} {\bibinfo {author} {\bibfnamefont {L.-K.}\ \bibnamefont {Lim}}, \bibinfo {author} {\bibfnamefont {J.-N.}\ \bibnamefont {Fuchs}}, \ and\ \bibinfo {author} {\bibfnamefont {G.}~\bibnamefont {Montambaux}},\ }\href {\doibase 10.1103/PhysRevLett.108.175303} {\bibfield  {journal} {\bibinfo  {journal} {Phys. Rev. Lett.}\ }\textbf {\bibinfo {volume} {108}},\ \bibinfo {pages} {175303} (\bibinfo {year} {2012})}\BibitemShut {NoStop}%
\bibitem [{\citenamefont {Franco De~Carvalho}\ and\ \citenamefont {Tavernelli}(2015)}]{franco_de_carvalho_nonadiabatic_2015}%
  \BibitemOpen
  \bibfield  {author} {\bibinfo {author} {\bibfnamefont {F.}~\bibnamefont {Franco De~Carvalho}}\ and\ \bibinfo {author} {\bibfnamefont {I.}~\bibnamefont {Tavernelli}},\ }\href {\doibase 10.1063/1.4936864} {\bibfield  {journal} {\bibinfo  {journal} {The Journal of Chemical Physics}\ }\textbf {\bibinfo {volume} {143}},\ \bibinfo {pages} {224105} (\bibinfo {year} {2015})}\BibitemShut {NoStop}%
\bibitem [{\citenamefont {Fuchs}\ \emph {et~al.}(2012)\citenamefont {Fuchs}, \citenamefont {Lim},\ and\ \citenamefont {Montambaux}}]{PhysRevA.86.063613}%
  \BibitemOpen
  \bibfield  {author} {\bibinfo {author} {\bibfnamefont {J.-N.}\ \bibnamefont {Fuchs}}, \bibinfo {author} {\bibfnamefont {L.-K.}\ \bibnamefont {Lim}}, \ and\ \bibinfo {author} {\bibfnamefont {G.}~\bibnamefont {Montambaux}},\ }\href {\doibase 10.1103/PhysRevA.86.063613} {\bibfield  {journal} {\bibinfo  {journal} {Phys. Rev. A}\ }\textbf {\bibinfo {volume} {86}},\ \bibinfo {pages} {063613} (\bibinfo {year} {2012})}\BibitemShut {NoStop}%
\bibitem [{\citenamefont {Krueckl}\ and\ \citenamefont {Richter}(2012)}]{krueckl_bloch-zener_2012}%
  \BibitemOpen
  \bibfield  {author} {\bibinfo {author} {\bibfnamefont {V.}~\bibnamefont {Krueckl}}\ and\ \bibinfo {author} {\bibfnamefont {K.}~\bibnamefont {Richter}},\ }\href {\doibase 10.1103/PhysRevB.85.115433} {\bibfield  {journal} {\bibinfo  {journal} {Phys. Rev. B}\ }\textbf {\bibinfo {volume} {85}},\ \bibinfo {pages} {115433} (\bibinfo {year} {2012})}\BibitemShut {NoStop}%
\bibitem [{\citenamefont {Kane}\ and\ \citenamefont {Mele}(2005)}]{Kane_2005}%
  \BibitemOpen
  \bibfield  {author} {\bibinfo {author} {\bibfnamefont {C.~L.}\ \bibnamefont {Kane}}\ and\ \bibinfo {author} {\bibfnamefont {E.~J.}\ \bibnamefont {Mele}},\ }\href {\doibase 10.1103/PhysRevLett.95.226801} {\bibfield  {journal} {\bibinfo  {journal} {Phys. Rev. Lett.}\ }\textbf {\bibinfo {volume} {95}},\ \bibinfo {pages} {226801} (\bibinfo {year} {2005})}\BibitemShut {NoStop}%
\bibitem [{\citenamefont {Kalesaki}\ \emph {et~al.}(2014)\citenamefont {Kalesaki}, \citenamefont {Delerue}, \citenamefont {Morais~Smith}, \citenamefont {Beugeling}, \citenamefont {Allan},\ and\ \citenamefont {Vanmaekelbergh}}]{PhysRevX.4.011010}%
  \BibitemOpen
  \bibfield  {author} {\bibinfo {author} {\bibfnamefont {E.}~\bibnamefont {Kalesaki}}, \bibinfo {author} {\bibfnamefont {C.}~\bibnamefont {Delerue}}, \bibinfo {author} {\bibfnamefont {C.}~\bibnamefont {Morais~Smith}}, \bibinfo {author} {\bibfnamefont {W.}~\bibnamefont {Beugeling}}, \bibinfo {author} {\bibfnamefont {G.}~\bibnamefont {Allan}}, \ and\ \bibinfo {author} {\bibfnamefont {D.}~\bibnamefont {Vanmaekelbergh}},\ }\href {\doibase 10.1103/PhysRevX.4.011010} {\bibfield  {journal} {\bibinfo  {journal} {Phys. Rev. X}\ }\textbf {\bibinfo {volume} {4}},\ \bibinfo {pages} {011010} (\bibinfo {year} {2014})}\BibitemShut {NoStop}%
\bibitem [{\citenamefont {Barreteau}\ \emph {et~al.}(2017)\citenamefont {Barreteau}, \citenamefont {Ducastelle},\ and\ \citenamefont {Mallah}}]{barreteau2017bird}%
  \BibitemOpen
  \bibfield  {author} {\bibinfo {author} {\bibfnamefont {C.}~\bibnamefont {Barreteau}}, \bibinfo {author} {\bibfnamefont {F.}~\bibnamefont {Ducastelle}}, \ and\ \bibinfo {author} {\bibfnamefont {T.}~\bibnamefont {Mallah}},\ }\href {\doibase 10.1088/1361-648X/aa8fec} {\bibfield  {journal} {\bibinfo  {journal} {J. Phys.: Cond. Mat.}\ }\textbf {\bibinfo {volume} {29}},\ \bibinfo {pages} {465302} (\bibinfo {year} {2017})}\BibitemShut {NoStop}%
\bibitem [{\citenamefont {Sharma}\ and\ \citenamefont {Banerjee}(2025)}]{sharma2025strain}%
  \BibitemOpen
  \bibfield  {author} {\bibinfo {author} {\bibfnamefont {S.}~\bibnamefont {Sharma}}\ and\ \bibinfo {author} {\bibfnamefont {A.~S.}\ \bibnamefont {Banerjee}},\ }\href {\doibase 10.48550/arXiv.2501.11783} {\bibfield  {journal} {\bibinfo  {journal} {arXiv preprint arXiv:2501.11783}\ } (\bibinfo {year} {2025}),\ 10.48550/arXiv.2501.11783}\BibitemShut {NoStop}%
\bibitem [{\citenamefont {Balents}\ \emph {et~al.}(2020)\citenamefont {Balents}, \citenamefont {Dean}, \citenamefont {Efetov},\ and\ \citenamefont {Young}}]{balents2020superconductivity}%
  \BibitemOpen
  \bibfield  {author} {\bibinfo {author} {\bibfnamefont {L.}~\bibnamefont {Balents}}, \bibinfo {author} {\bibfnamefont {C.~R.}\ \bibnamefont {Dean}}, \bibinfo {author} {\bibfnamefont {D.~K.}\ \bibnamefont {Efetov}}, \ and\ \bibinfo {author} {\bibfnamefont {A.~F.}\ \bibnamefont {Young}},\ }\href {\doibase 10.1038/s41567-020-0906-9} {\bibfield  {journal} {\bibinfo  {journal} {Nature Physics}\ }\textbf {\bibinfo {volume} {16}},\ \bibinfo {pages} {725} (\bibinfo {year} {2020})}\BibitemShut {NoStop}%
\bibitem [{\citenamefont {Iglovikov}\ \emph {et~al.}(2014)\citenamefont {Iglovikov}, \citenamefont {Hébert}, \citenamefont {Grémaud}, \citenamefont {Batrouni},\ and\ \citenamefont {Scalettar}}]{iglovikov2014superconducting}%
  \BibitemOpen
  \bibfield  {author} {\bibinfo {author} {\bibfnamefont {V.~I.}\ \bibnamefont {Iglovikov}}, \bibinfo {author} {\bibfnamefont {F.}~\bibnamefont {Hébert}}, \bibinfo {author} {\bibfnamefont {B.}~\bibnamefont {Grémaud}}, \bibinfo {author} {\bibfnamefont {G.~G.}\ \bibnamefont {Batrouni}}, \ and\ \bibinfo {author} {\bibfnamefont {R.~T.}\ \bibnamefont {Scalettar}},\ }\href {\doibase 10.1103/PhysRevB.90.094506} {\bibfield  {journal} {\bibinfo  {journal} {Phys. Rev. B}\ }\textbf {\bibinfo {volume} {90}},\ \bibinfo {pages} {094506} (\bibinfo {year} {2014})}\BibitemShut {NoStop}%
\bibitem [{\citenamefont {Goldman}\ \emph {et~al.}(2012)\citenamefont {Goldman}, \citenamefont {Beugnon},\ and\ \citenamefont {Gerbier}}]{Aidelsburger_2013}%
  \BibitemOpen
  \bibfield  {author} {\bibinfo {author} {\bibfnamefont {N.}~\bibnamefont {Goldman}}, \bibinfo {author} {\bibfnamefont {J.}~\bibnamefont {Beugnon}}, \ and\ \bibinfo {author} {\bibfnamefont {F.}~\bibnamefont {Gerbier}},\ }\href {\doibase 10.1103/PhysRevLett.108.255303} {\bibfield  {journal} {\bibinfo  {journal} {Phys. Rev. Lett.}\ }\textbf {\bibinfo {volume} {108}},\ \bibinfo {pages} {255303} (\bibinfo {year} {2012})}\BibitemShut {NoStop}%
\bibitem [{\citenamefont {Liu}\ \emph {et~al.}(2014)\citenamefont {Liu}, \citenamefont {Liu},\ and\ \citenamefont {Wu}}]{Liu2014FlatBands}%
  \BibitemOpen
  \bibfield  {author} {\bibinfo {author} {\bibfnamefont {Z.}~\bibnamefont {Liu}}, \bibinfo {author} {\bibfnamefont {F.}~\bibnamefont {Liu}}, \ and\ \bibinfo {author} {\bibfnamefont {Y.-S.}\ \bibnamefont {Wu}},\ }\href {\doibase 10.1088/1674-1056/23/7/077308} {\bibfield  {journal} {\bibinfo  {journal} {Chinese Physics B}\ }\textbf {\bibinfo {volume} {23}},\ \bibinfo {pages} {077308} (\bibinfo {year} {2014})}\BibitemShut {NoStop}%
\bibitem [{\citenamefont {Khomeriki}\ and\ \citenamefont {Flach}(2016)}]{PhysRevLett.116.245301}%
  \BibitemOpen
  \bibfield  {author} {\bibinfo {author} {\bibfnamefont {R.}~\bibnamefont {Khomeriki}}\ and\ \bibinfo {author} {\bibfnamefont {S.}~\bibnamefont {Flach}},\ }\href {\doibase 10.1103/PhysRevLett.116.245301} {\bibfield  {journal} {\bibinfo  {journal} {Phys. Rev. Lett.}\ }\textbf {\bibinfo {volume} {116}},\ \bibinfo {pages} {245301} (\bibinfo {year} {2016})}\BibitemShut {NoStop}%
\bibitem [{\citenamefont {Ye}\ and\ \citenamefont {Lai}(2023)}]{ye_irregular_2023}%
  \BibitemOpen
  \bibfield  {author} {\bibinfo {author} {\bibfnamefont {L.-L.}\ \bibnamefont {Ye}}\ and\ \bibinfo {author} {\bibfnamefont {Y.-C.}\ \bibnamefont {Lai}},\ }\href {\doibase 10.1103/PhysRevB.107.165422} {\bibfield  {journal} {\bibinfo  {journal} {Phys. Rev. B}\ }\textbf {\bibinfo {volume} {107}},\ \bibinfo {pages} {165422} (\bibinfo {year} {2023})}\BibitemShut {NoStop}%
\bibitem [{\citenamefont {Lim}\ \emph {et~al.}(2020)\citenamefont {Lim}, \citenamefont {Fuchs}, \citenamefont {Piéchon},\ and\ \citenamefont {Montambaux}}]{lim_dirac_2020}%
  \BibitemOpen
  \bibfield  {author} {\bibinfo {author} {\bibfnamefont {L.-K.}\ \bibnamefont {Lim}}, \bibinfo {author} {\bibfnamefont {J.-N.}\ \bibnamefont {Fuchs}}, \bibinfo {author} {\bibfnamefont {F.}~\bibnamefont {Piéchon}}, \ and\ \bibinfo {author} {\bibfnamefont {G.}~\bibnamefont {Montambaux}},\ }\href {\doibase 10.1103/PhysRevB.101.045131} {\bibfield  {journal} {\bibinfo  {journal} {Phys. Rev. B}\ }\textbf {\bibinfo {volume} {101}},\ \bibinfo {pages} {045131} (\bibinfo {year} {2020})}\BibitemShut {NoStop}%
\bibitem [{\citenamefont {Guo}\ and\ \citenamefont {Franz}(2009)}]{PhysRevB.80.113102}%
  \BibitemOpen
  \bibfield  {author} {\bibinfo {author} {\bibfnamefont {H.-M.}\ \bibnamefont {Guo}}\ and\ \bibinfo {author} {\bibfnamefont {M.}~\bibnamefont {Franz}},\ }\href {\doibase 10.1103/PhysRevB.80.113102} {\bibfield  {journal} {\bibinfo  {journal} {Phys. Rev. B}\ }\textbf {\bibinfo {volume} {80}},\ \bibinfo {pages} {113102} (\bibinfo {year} {2009})}\BibitemShut {NoStop}%
\bibitem [{\citenamefont {Bergholtz}\ and\ \citenamefont {Liu}(2013)}]{BERGHOLTZ_2013}%
  \BibitemOpen
  \bibfield  {author} {\bibinfo {author} {\bibfnamefont {E.~J.}\ \bibnamefont {Bergholtz}}\ and\ \bibinfo {author} {\bibfnamefont {Z.}~\bibnamefont {Liu}},\ }\href {\doibase 10.1142/s021797921330017x} {\bibfield  {journal} {\bibinfo  {journal} {International Journal of Modern Physics B}\ }\textbf {\bibinfo {volume} {27}},\ \bibinfo {pages} {1330017} (\bibinfo {year} {2013})}\BibitemShut {NoStop}%
\bibitem [{\citenamefont {Yin}\ \emph {et~al.}(2022)\citenamefont {Yin}, \citenamefont {Lian},\ and\ \citenamefont {Hasan}}]{Yin_2022}%
  \BibitemOpen
  \bibfield  {author} {\bibinfo {author} {\bibfnamefont {J.-X.}\ \bibnamefont {Yin}}, \bibinfo {author} {\bibfnamefont {B.}~\bibnamefont {Lian}}, \ and\ \bibinfo {author} {\bibfnamefont {M.~Z.}\ \bibnamefont {Hasan}},\ }\href {\doibase 10.1038/s41586-022-05516-0} {\bibfield  {journal} {\bibinfo  {journal} {Nature}\ }\textbf {\bibinfo {volume} {612}},\ \bibinfo {pages} {647–657} (\bibinfo {year} {2022})}\BibitemShut {NoStop}%
\bibitem [{\citenamefont {Leykam}\ \emph {et~al.}(2018)\citenamefont {Leykam}, \citenamefont {Andreanov},\ and\ \citenamefont {Flach}}]{Leykam_2018}%
  \BibitemOpen
  \bibfield  {author} {\bibinfo {author} {\bibfnamefont {D.}~\bibnamefont {Leykam}}, \bibinfo {author} {\bibfnamefont {A.}~\bibnamefont {Andreanov}}, \ and\ \bibinfo {author} {\bibfnamefont {S.}~\bibnamefont {Flach}},\ }\href {\doibase 10.1080/23746149.2018.1473052} {\bibfield  {journal} {\bibinfo  {journal} {Advances in Physics: X}\ }\textbf {\bibinfo {volume} {3}},\ \bibinfo {pages} {1473052} (\bibinfo {year} {2018})}\BibitemShut {NoStop}%
\bibitem [{\citenamefont {Jiang}\ \emph {et~al.}(2019)\citenamefont {Jiang}, \citenamefont {Kang}, \citenamefont {Huang}, \citenamefont {Xu}, \citenamefont {Low},\ and\ \citenamefont {Liu}}]{jiang_topological_2019}%
  \BibitemOpen
  \bibfield  {author} {\bibinfo {author} {\bibfnamefont {W.}~\bibnamefont {Jiang}}, \bibinfo {author} {\bibfnamefont {M.}~\bibnamefont {Kang}}, \bibinfo {author} {\bibfnamefont {H.}~\bibnamefont {Huang}}, \bibinfo {author} {\bibfnamefont {H.}~\bibnamefont {Xu}}, \bibinfo {author} {\bibfnamefont {T.}~\bibnamefont {Low}}, \ and\ \bibinfo {author} {\bibfnamefont {F.}~\bibnamefont {Liu}},\ }\href {\doibase 10.1103/PhysRevB.99.125131} {\bibfield  {journal} {\bibinfo  {journal} {Phys. Rev. B}\ }\textbf {\bibinfo {volume} {99}},\ \bibinfo {pages} {125131} (\bibinfo {year} {2019})}\BibitemShut {NoStop}%
\bibitem [{\citenamefont {Ando}(2013)}]{Ando_2005}%
  \BibitemOpen
  \bibfield  {author} {\bibinfo {author} {\bibfnamefont {Y.}~\bibnamefont {Ando}},\ }\href {\doibase 10.7566/JPSJ.82.102001} {\bibfield  {journal} {\bibinfo  {journal} {Journal of the Physical Society of Japan}\ }\textbf {\bibinfo {volume} {82}},\ \bibinfo {pages} {102001} (\bibinfo {year} {2013})}\BibitemShut {NoStop}%
\bibitem [{\citenamefont {Cristiani}\ \emph {et~al.}(2002)\citenamefont {Cristiani}, \citenamefont {Morsch}, \citenamefont {Müller}, \citenamefont {Ciampini},\ and\ \citenamefont {Arimondo}}]{BEC_2006}%
  \BibitemOpen
  \bibfield  {author} {\bibinfo {author} {\bibfnamefont {M.}~\bibnamefont {Cristiani}}, \bibinfo {author} {\bibfnamefont {O.}~\bibnamefont {Morsch}}, \bibinfo {author} {\bibfnamefont {J.~H.}\ \bibnamefont {Müller}}, \bibinfo {author} {\bibfnamefont {D.}~\bibnamefont {Ciampini}}, \ and\ \bibinfo {author} {\bibfnamefont {E.}~\bibnamefont {Arimondo}},\ }\href {\doibase 10.1103/PhysRevA.65.063612} {\bibfield  {journal} {\bibinfo  {journal} {Phys. Rev. A}\ }\textbf {\bibinfo {volume} {65}},\ \bibinfo {pages} {063612} (\bibinfo {year} {2002})}\BibitemShut {NoStop}%
\bibitem [{\citenamefont {Rosenstein}\ \emph {et~al.}(2010)\citenamefont {Rosenstein}, \citenamefont {Lewkowicz}, \citenamefont {Kao},\ and\ \citenamefont {Korniyenko}}]{rosenstein_ballistic_2010}%
  \BibitemOpen
  \bibfield  {author} {\bibinfo {author} {\bibfnamefont {B.}~\bibnamefont {Rosenstein}}, \bibinfo {author} {\bibfnamefont {M.}~\bibnamefont {Lewkowicz}}, \bibinfo {author} {\bibfnamefont {H.~C.}\ \bibnamefont {Kao}}, \ and\ \bibinfo {author} {\bibfnamefont {Y.}~\bibnamefont {Korniyenko}},\ }\href {\doibase 10.1103/PhysRevB.81.041416} {\bibfield  {journal} {\bibinfo  {journal} {Phys. Rev. B}\ }\textbf {\bibinfo {volume} {81}},\ \bibinfo {pages} {041416} (\bibinfo {year} {2010})}\BibitemShut {NoStop}%
\bibitem [{\citenamefont {Cancès}\ \emph {et~al.}(2021)\citenamefont {Cancès}, \citenamefont {Fermanian Kammerer}, \citenamefont {Levitt},\ and\ \citenamefont {Siraj-Dine}}]{cances_coherent_2021}%
  \BibitemOpen
  \bibfield  {author} {\bibinfo {author} {\bibfnamefont {E.}~\bibnamefont {Cancès}}, \bibinfo {author} {\bibfnamefont {C.}~\bibnamefont {Fermanian Kammerer}}, \bibinfo {author} {\bibfnamefont {A.}~\bibnamefont {Levitt}}, \ and\ \bibinfo {author} {\bibfnamefont {S.}~\bibnamefont {Siraj-Dine}},\ }\href {\doibase 10.1007/s00023-021-01026-3} {\bibfield  {journal} {\bibinfo  {journal} {Ann. Henri Poincaré}\ }\textbf {\bibinfo {volume} {22}},\ \bibinfo {pages} {2643} (\bibinfo {year} {2021})}\BibitemShut {NoStop}%
\bibitem [{\citenamefont {Monkhorst}\ and\ \citenamefont {Pack}(1976)}]{PhysRevB.13.5188}%
  \BibitemOpen
  \bibfield  {author} {\bibinfo {author} {\bibfnamefont {H.~J.}\ \bibnamefont {Monkhorst}}\ and\ \bibinfo {author} {\bibfnamefont {J.~D.}\ \bibnamefont {Pack}},\ }\href {\doibase 10.1103/PhysRevB.13.5188} {\bibfield  {journal} {\bibinfo  {journal} {Phys. Rev. B}\ }\textbf {\bibinfo {volume} {13}},\ \bibinfo {pages} {5188} (\bibinfo {year} {1976})}\BibitemShut {NoStop}%
\bibitem [{\citenamefont {Yoshida}(1990)}]{yoshida_construction_1990}%
  \BibitemOpen
  \bibfield  {author} {\bibinfo {author} {\bibfnamefont {H.}~\bibnamefont {Yoshida}},\ }\href {\doibase 10.1016/0375-9601(90)90092-3} {\bibfield  {journal} {\bibinfo  {journal} {Physics Letters A}\ }\textbf {\bibinfo {volume} {150}},\ \bibinfo {pages} {262} (\bibinfo {year} {1990})}\BibitemShut {NoStop}%
\bibitem [{\citenamefont {Fehlberg}(1969)}]{Fehlberg1969LoworderCR}%
  \BibitemOpen
  \bibfield  {author} {\bibinfo {author} {\bibfnamefont {E.}~\bibnamefont {Fehlberg}},\ }\href {https://ntrs.nasa.gov/citations/19690021375} {\emph {\bibinfo {title} {Low-order classical Runge-Kutta formulas with stepsize control and their application to some heat transfer problems}}},\ \bibinfo {type} {Tech. Rep.}\ \bibinfo {number} {NASA TR R-315}\ (\bibinfo  {institution} {NASA},\ \bibinfo {address} {Washington, D.C.},\ \bibinfo {year} {1969})\BibitemShut {NoStop}%
\bibitem [{\citenamefont {Fahimniya}\ \emph {et~al.}(2021)\citenamefont {Fahimniya}, \citenamefont {Dong}, \citenamefont {Kiselev},\ and\ \citenamefont {Levitov}}]{fahimniya_synchronizing_2021}%
  \BibitemOpen
  \bibfield  {author} {\bibinfo {author} {\bibfnamefont {A.}~\bibnamefont {Fahimniya}}, \bibinfo {author} {\bibfnamefont {Z.}~\bibnamefont {Dong}}, \bibinfo {author} {\bibfnamefont {E.~I.}\ \bibnamefont {Kiselev}}, \ and\ \bibinfo {author} {\bibfnamefont {L.}~\bibnamefont {Levitov}},\ }\href {\doibase 10.1103/PhysRevLett.126.256803} {\bibfield  {journal} {\bibinfo  {journal} {Phys. Rev. Lett.}\ }\textbf {\bibinfo {volume} {126}},\ \bibinfo {pages} {256803} (\bibinfo {year} {2021})}\BibitemShut {NoStop}%
\bibitem [{\citenamefont {Sakurai}\ and\ \citenamefont {Napolitano}(2020)}]{sakurai1994modern}%
  \BibitemOpen
  \bibfield  {author} {\bibinfo {author} {\bibfnamefont {J.~J.}\ \bibnamefont {Sakurai}}\ and\ \bibinfo {author} {\bibfnamefont {J.}~\bibnamefont {Napolitano}},\ }\href {\doibase 10.1017/9781108587280} {\emph {\bibinfo {title} {Modern Quantum Mechanics}}},\ \bibinfo {edition} {3rd}\ ed.\ (\bibinfo  {publisher} {Cambridge University Press},\ \bibinfo {year} {2020})\BibitemShut {NoStop}%
\bibitem [{\citenamefont {Dai}\ \emph {et~al.}(2019)\citenamefont {Dai}, \citenamefont {Liu},\ and\ \citenamefont {Zhang}}]{dai_strain_2019}%
  \BibitemOpen
  \bibfield  {author} {\bibinfo {author} {\bibfnamefont {Z.}~\bibnamefont {Dai}}, \bibinfo {author} {\bibfnamefont {L.}~\bibnamefont {Liu}}, \ and\ \bibinfo {author} {\bibfnamefont {Z.}~\bibnamefont {Zhang}},\ }\href {\doibase 10.1002/adma.201805417} {\bibfield  {journal} {\bibinfo  {journal} {Advanced Materials}\ }\textbf {\bibinfo {volume} {31}},\ \bibinfo {pages} {1805417} (\bibinfo {year} {2019})}\BibitemShut {NoStop}%
\bibitem [{\citenamefont {Lozier}(2003)}]{lozier2003nist}%
  \BibitemOpen
  \bibfield  {author} {\bibinfo {author} {\bibfnamefont {D.~W.}\ \bibnamefont {Lozier}},\ }\href {https://dlmf.nist.gov/} {\bibfield  {journal} {\bibinfo  {journal} {Annals of Mathematics and Artificial Intelligence}\ }\textbf {\bibinfo {volume} {38}},\ \bibinfo {pages} {105} (\bibinfo {year} {2003})}\BibitemShut {NoStop}%
\end{thebibliography}%

\section{Appendices} 
The fast time evolution of three-level quantum mechanical systems, especially those with band openings at crossing points, can be effectively described using the non-adiabatic approximation within the first-order expansion of the tight-binding Hamiltonian. In Kagome and Lieb lattices, this approach captures the influence of external perturbations on quantum state transitions, providing insights into non-equilibrium dynamics, Landau-Zener transitions, and band structure modifications driven by strain or external fields. To fundamentally understand the impact of Landau-Zener transitions (LZT) on each energy level during the non-adiabatic process, we solve the time-dependent Schrödinger equation (TDSE) for these three-level systems. This enables us to directly observe the temporal evolution of an initial wave packet, examine transition probabilities, and analyze potential energy level transitions driven by external perturbations such as electric fields or strain.

\subsection{Kagome Lattice}
Given the Hamiltonian for the Kagome lattice in Equation~\ref{Kagome_expansion}, the time-dependent Schrödinger equation leads to two coupled equations for the components of $\phi(t)$:
\setlength{\jot}{0pt}
\begin{align}
    i \frac{d\phi_1}{dt} &= \epsilon \beta^2 t^2 \phi_1 + g \beta t \phi_2, \\
    \vspace{-5pt}
    i \frac{d\phi_2}{dt} &= g \beta t \phi_1 + \frac{g^2}{\epsilon}\phi_2.
\end{align}
We can eliminate $\phi_2$ using its expression in terms of  $\phi_1$. After some algebra we find
\begin{equation}
t \frac{d^2 \phi_1}{dt^2} + \left( \frac{i g^2}{\epsilon} t + i \epsilon \beta^2 t^3 - 1 \right) \frac{d \phi_1}{dt} + i \epsilon \beta^2 t^2 \phi_1 = 0.
\end{equation}

\subsection{Lieb Lattice}
Given the Hamiltonian for the Lieb lattice in Equation \ref{expansion_lieb_simply}, the time-dependent Schrödinger equation leads to three coupled bands with a set of equations of $\phi(t)$:
\setlength{\jot}{0pt}
\begin{align}\label{eqn:Lieb-system}
    i \frac{d\phi_1}{dt} &= \epsilon t \phi_1 + g \phi_3, \\
    i \frac{d\phi_2}{dt} &= - \epsilon t \phi_2 + g \phi_3, \\
    i \frac{d\phi_3}{dt} &= g \phi_1 + g \phi_2.
\end{align}
We can use these equations to express $\phi_2$ and $\phi_3$ in terms of $\phi_1$. After some algebra we find
\begin{equation}
\label{phi_1}
\frac{d^3 \phi_1}{dt^3} + \left( 2 i\epsilon+ 2 g^2 + \epsilon^2 t^2 \right) \frac{d \phi_1}{dt} + \epsilon^2 \phi_1 = 0.
\end{equation}
It is possible to obtain analytical solutions to the above equation in terms of parabolic cylinder functions \cite{lozier2003nist}, which we note below for the sake of completeness:
\begin{align}
&\phi_1 = C_{1} \left[D \left(-\frac{i g^2}{2 \epsilon}, i^{1/2} \sqrt{\epsilon} \, t\right)\right]^2  \nonumber \\
& + C_{2} \left[D \left(-\frac{i g^2}{2 \epsilon}, i^{1/2} \sqrt{\epsilon} \, t\right)\right] 
\left[D\left(\frac{-2 \epsilon + i g^2}{2 \epsilon}, i^{3/2} \sqrt{\epsilon} \, t\right)\right] \nonumber \\
& + C_{3} \left[D \left(\frac{-2 \epsilon + i g^2}{2 \epsilon}, i^{3/2} \sqrt{\epsilon} \, t\right)\right]^2.
\label{solution_1_parameter}
\end{align}
Thereafter, we can use equations (\ref{eqn:Lieb-system}) to obtain expressions for $\phi_2$ and $\phi_3$ as well:
\begin{equation}
\begin{aligned}
& \phi_2 = -\frac{1}{g} \sqrt{a} \Bigg( 2 \sqrt{a} t C_1 
D \left(-\frac{i g^2}{2 a}, (-i)^{1/2} \sqrt{a} t \right)^2 
+ i^{1/2} \bigg( -2 C_3 
D \left(\frac{i g^2}{2 a}, i^{3/2} \sqrt{a} t \right) \\
& \quad + i C_2 D \left(1 - \frac{i g^2}{2 a}, i^{1/2} \sqrt{a} t \right) 
\bigg) D \left(-1 + \frac{i g^2}{2 a}, i^{3/2} \sqrt{a} t \right)\\ 
& \quad + D \left(-\frac{i g^2}{2 a}, i^{1/2} \sqrt{a} t \right) 
\bigg( -i^{1/2} C_2 
D \left(\frac{i g^2}{2 a}, i^{3/2} \sqrt{a} t \right) 
+ 2 i^{3/2} C_1 
D \left(1 - \frac{i g^2}{2 a}, i^{1/2} \sqrt{a} t \right) \\
& \quad + \sqrt{a} t C_2 
D \left(-1 + \frac{i g^2}{2 a}, i^{3/2} \sqrt{a} t \right) 
\bigg) \Bigg),
\end{aligned}
\label{solution_2_parameter}
\end{equation}
and
\begin{equation}
\begin{aligned}
    \phi_3 &=\frac{1}{g} \left( I \frac{\partial}{\partial t} \left( 
    -\frac{1}{g} \sqrt{a} \Bigg( 
    2 \sqrt{a} t C_1 
    D \left(-\frac{I g^2}{2 a}, \left(-1\right)^{1/4} \sqrt{a} t \right)^2 \right.\right. \\ 
    &\quad + (-1)^{1/4} \bigg( 
    -2 C_3 
    D \left(\frac{I g^2}{2 a}, \left(-1\right)^{3/4} \sqrt{a} t \right) 
    + I C_2 D \left(1 - \frac{I g^2}{2 a}, \left(-1\right)^{1/4} \sqrt{a} t \right) 
    \bigg) \\ 
    &\quad \times D \left(-1 + \frac{I g^2}{2 a}, \left(-1\right)^{3/4} \sqrt{a} t \right) 
    + D \left(-\frac{I g^2}{2 a}, \left(-1\right)^{1/4} \sqrt{a} t \right) \bigg( 
    -(-1)^{1/4} C_2 
    D \left(\frac{I g^2}{2 a}, \left(-1\right)^{3/4} \sqrt{a} t \right) \\ 
    &\quad + 2 (-1)^{3/4} C_1 
    D \left(1 - \frac{I g^2}{2 a}, \left(-1\right)^{1/4} \sqrt{a} t \right) 
    + \sqrt{a} t C_2 
    D \left(-1 + \frac{I g^2}{2 a}, \left(-1\right)^{3/4} \sqrt{a} t \right) 
    \bigg) \Bigg) ) \\ 
    &\quad - g \Bigg( 
    C_1 D \left(-\frac{I g^2}{2 a}, \left(-1\right)^{1/4} \sqrt{a} t \right)^2 
    + C_2 
    D \left(-\frac{I g^2}{2 a}, \left(-1\right)^{1/4} \sqrt{a} t \right) 
    D \left(-1 + \frac{I g^2}{2 a}, \left(-1\right)^{3/4} \sqrt{a} t \right) \\ 
    &\quad + C_3 
    D \left(-1 + \frac{I g^2}{2 a}, \left(-1\right)^{3/4} \sqrt{a} t \right)^2 
    \Bigg).
\end{aligned}
\label{solution_3_parameter}
\end{equation}
In the remote past ($t \to -\infty$), we assume the system is prepared in its ground state, so only the lowest-energy level is initially occupied. Accordingly, the state vector is taken to have components $\phi_1(t_0) = 1$, $\phi_2(t_0) = 0$, and $\phi_3(t_0) = 0$.

These constraints and the relations between $\phi_1$ with respect to $\phi_2$ and $\phi_3$ allow us to determine the initial values of $\phi_1(t)$ and its first and second derivatives, given by: $\dot{\phi}_1(t_0) = -i\epsilon t_0$, and $\ddot{\phi}_1(t_0) = -i\epsilon - (\epsilon t_0)^2 + g^2 $

With these conditions, the coefficients $C_1$, $C_2$ and $C_3$ in Equation \ref{phi_1} can determine a closed-form analytical solution contain special function, which we verified using a MATLAB package. However, the resulting expressions are complicated and do not provide direct physical intuition, so we do not present them here. For our purposes, it is sufficient to note that the solution can be constructed in principle, and we focus on the qualitative behavior they capture.


\end{document}